\documentclass[fleqn]{article} 
\usepackage[T1]{fontenc}
\usepackage[utf8]{inputenc}
\usepackage[a4paper]{geometry}
\usepackage{float}
\usepackage{textcomp}
\usepackage{amstext}
\usepackage{amssymb}
\usepackage{stmaryrd}
\usepackage{graphicx}
\usepackage{subscript}
\usepackage{placeins}
\usepackage{amsmath} 
\usepackage{authblk}
\usepackage{hyperref} 
\usepackage{placeins} 
\usepackage{tikz-cd}

\makeatletter

\providecommand{\tabularnewline}{\\}

\floatstyle{ruled}
\newfloat{algorithm}{tbp}{loa}
\providecommand{\algorithmname}{Algorithm}
\floatname{algorithm}{\protect\algorithmname}

\makeatother

\usepackage{babel}
\begin{document}
\title{IBAC Mathematics and Mechanics\\ The Case for `Integer Based Access Control'\\ \Large{Data Security in the Age of AI and AI Automation}}

\author{Mark Stocks\\
   Email: mark.stocks@canberra.edu.au \\
   ORCID: 0000-0001-9155-4276}
\affil{Canberra University}
\date{\today}

\maketitle

\begin{abstract}
\setlength{\parindent}{0pt}
\noindent Current methods for data access control, especially in regards to AI and AI automation face unique challenges in ensuring appropriate data access. We introduce Integer-Based Access Control (IBAC), addressing the limitations of Role-Based Access Control (RBAC) and Attribute-Based Access Control (ABAC). IBAC’s mathematical foundations enable its application to relational and NoSQL databases and document authorisation. We demonstrate IBAC’s suitability for filtering relational database row-level information as well as  AI prompt search results, particularly in restricting AI/NLP access based on separation of duty, supporting both need-to-know and need-to-share data restrictions. \par

 \medskip

IBAC uses security tokens, integers that represent aggregated security attributes. These tokens maintain orthogonality across encoded attributes, but are stored as integers for fast real-time vector comparison and efficient dominance testing. This mechanism allows high-speed row-level result filtering, ensuring unauthorised records are excluded before results reach the requester. \par

\medskip

We extend the Bell-LaPadula model by incorporating a 'process constraint', overcoming RBAC/ABAC limitations with reduced complexity, increased flexibility, and enhanced performance in data filtering. Our theorems demonstrate the extended Dominance relationship, facilitating rapid federated authorisation across diverse databases and file systems. \par

\medskip

This work reaffirms the practical strength of the Bell-LaPadula model in data security through (1) our mathematical extension, (2) a novel IBAC security attribute encoding scheme, and (3) a simplified dominance testing mechanism of security tokens without decoding. \par
\end{abstract}

\section{Background}
In today’s zero-trust cyber environment, data security is paramount, as both external and internal threats continually challenge organisations. Traditional security measures often fail, especially against internal threats that emerge once an outsider breaches initial defences. This paper introduces Integer-Based Access Control (IBAC) to enhance data security, particularly in AI and automation contexts. \par

\medskip

Historically, business and IT leaders have focused on perimeter defences, guided by industry advice. However, the increasing frequency of data breaches shows that this approach is insufficient. Data breaches typically follow a two-step process: an external attack breaches the perimeter (often through phishing), and the intruder then acts as an internal threat, similar to a rogue employee. \par

\medskip

Although external attacks are concerning, the real risk arises when external attacks evolve into insider threats. Many organisations overlook this two-step process, finding that the internal threat is too complex to address with the known complexities of exiting access controls. \par

\medskip

Access control involves two main processes: authentication and authorisation. Authentication verifies the user's identity through credentials, which form the first layer of security. Authorisation determines what authenticated users can access based on predefined policies such as RBAC, ABAC, or MAC. Current practices rely heavily on Active Directory and application functionality, which often leave data vulnerable. \par

\medskip

This paper advocates for stronger internal data authorisation controls. We propose IBAC as a robust and simpler method to enforce zero trust at the data level, ensuring data protection even if authentication or application security is compromised. IBAC provides an independent layer of defence, enforcing need-to-know and need-to-share principles for data access. \par

\medskip

Our focus is on inward-focused authorisation controls, critiquing existing mechanisms (RBAC, ABAC, CapBAC), and presenting IBAC as a superior alternative for secure, efficient data access management. By implementing IBAC, we demonstrate a significant improvement in data authorisation, reducing complexity, and enhancing security within organisations. \par

\medskip

\subsection{Role-Based Access Control (RBAC) limitations}  RBAC assigns permissions based on roles and is widely used, but has significant constraints. It often provides excessive access to senior technical personnel and does not prevent data exfiltration through backdoors or compromised accounts. RBAC is effective for role-based access control, but was never designed for fine-grained data access control. We will see that IBAC extends RBAC, creating a hybrid approach. IBAC provides a more detailed and fine-grained access control, that limits access to senior technical personnel and is designed to close back doors and stop data exfiltration.  \par

\medskip 

\subsection{Active Directory (AD) limitations} AD extends RBAC but still inherits its fundamental weaknesses, particularly in managing granular data access and integrating with modern security needs. \par

\medskip 

\subsection {Attribute-Based Access Control (ABAC) limitations} ABAC \cite{key-8, key-9} offers more granular control by evaluating multiple attributes. IBAC was developed to overcome the limitation of ABAC, so it deserves a little more attention to understand limitations. ABAC implementation is complex, performance-intensive, and lacks standardisation, making it difficult for large organisations to adopt and maintain effectively. Limitations include: \par

\begin{itemize}
    \item High Initial Effort: Defining attributes, establishing policies, and integrating with existing systems require significant upfront work.
    \item Maintenance Complexity: Developing and maintaining attribute schemas and policies is resource-intensive.
    \item Performance Bottlenecks: Real-time evaluation of access requests can be computationally intensive, leading to performance issues.
    \item Scalability Issues: The system may struggle to handle high volumes of access requests and complex evaluations.
    \item Standardization and Compatibility: Lack of widely accepted standards can cause compatibility issues between different systems and implementations.
    \item Vendor Integration Challenges: Integrating ABAC solutions from different vendors can be difficult.
    \item Specialized Knowledge Required: Administrators and policy creators need specialized skills to manage ABAC systems effectively.
    \item Policy Validation: Extensive testing and validation are needed to ensure policies do not result in unintended access.
    \item Legacy System Integration: Integrating with legacy systems may require substantial modifications.
    \item Hybrid Environment Complexity: Managing access control in environments combining ABAC with other models like RBAC adds complexity.
\end{itemize}

\medskip 

\subsection{Capability-Based Access Control (CapBAC) limitations} CapBAC \cite{key-10, key-11} has a concept of security tokens, similar to IBAC, so it deserves a little more attention as well.  CapBAC provides a flexible and decentralised way to manage access control using tokens that specify access rights. However, CapBAC also has numerous limitations, here are some of the limitations of CapBAC: \par

\medskip 

\begin{itemize}
    \item Life-cycle Management: Managing the life-cycle of capability tokens (issuance, renewal, revocation, and expiration) can be complex, especially in large-scale systems.
    \item Token distribution: Ensuring that tokens are securely distributed and stored is challenging, as tokens can be susceptible to theft or misuse if not properly protected.
    \item Token explosion: In systems with a vast number of resources and users, the number of tokens that need to be managed can become very large, leading to potential scalability issues.
    \item Token Security Risks: If a capability token is stolen or intercepted, it can be used by an unauthorized party to access resources. This makes the secure handling and transmission of tokens critical.
    \item Security Handing Overheads: Ensuring the integrity and authenticity of tokens often requires cryptographic measures, which can introduce additional overhead and performance degradation.
    \item Complicated Governance: While CapBAC allows for fine-grained access control, defining and managing the granularity of capabilities can be cumbersome.
    \item Unauthorised Access Risks: Delegating capabilities must be carefully controlled to prevent unauthorized access through improper delegation.
    \item Integrating CapBAC: Integrating CapBAC with existing access control systems and policies can be challenging. It requires a shift in how access control is conceptualized and implemented.
    \item Cognitive load: Users and administrators must understand and correctly handle capability tokens, which can increase the cognitive load and potential for errors.
\end{itemize}

\medskip 
\subsection{Integer Based Access Control (IBAC) and it's Trusted-System linage}

Against this backdrop, we discuss and look at the details of integer-based attribute control (IBAC) compared to RBAC, ABAC, and CapBAC, and we demonstrate that IBAC addresses the major limitations of the other methods above.  \par

\medskip 

Historically Trusted Systems, particularly Multilevel Security (MLS) systems, were based on the Bell-LaPadula and Biba mathematical models. Eventually, attribute-based access control (ABAC) and capability-based access control (CapBAC) were developed to address perceived limitations and inflexibilities of traditional Trusted Systems based on these earlier models. \par

\medskip 

We argue that moving away from the mathematically sound trusted Bell-LaPadula and Biba models to build access control systems like ABAC and CapBAC, which lack a solid mathematical foundation, was a mistake. Instead, extending the Bell-LaPadula and Biba mathematics to enhance robustness and introduce flexibility should have been the focus. \par

\medskip

ABAC and CapBAC are based on testing and verification. The absence of a mathematical basis for the ABAC and CapBAC models means that each time a policy is defined, there is a risk of inadvertently introducing a backdoor or security vulnerability in the ABAC/CapBAC framework. However, this is not the case for the Bell-LaPadula and Biba trust models. \par

\medskip 

The access control models post Trusted-Systems Bell-LaPadula and Biba that we have discussed, i.e., ABAC, CapBAC, have had little impact todate in addressing global challenges of managing the growing wave of cybersecurity threats that result in data exfiltration. We observe a steady rise in data breaches, prompting governments to respond with continuous legislation. \par

\medskip 

We would argue the original trusted system Bell-Lapadual and Biba models was sound, and at the heart of the model, more than capable in solving the authorisation problem, but the implementation of the mathematical model was flawed. Trusted systems have been around since the 1970s with little impact on the global problems that we face today in terms of managing the ever-increasing wave of cyber security threats. Over the years, a lot of \textquotedblleft off-the-cuff\textquotedblright{} criticism has been leveled at the trusted system and MLS approach, specifically how inflexible and nonperforming the method is. We disagree with that assessment and show in this paper that the issue with the trusted system, specifically Bell-Lapadual and Biba, has not been the mathematical model, but the traditional implementation using bit vectors. The bit-vector implementation might be easy to deploy, but it is easy to hack, it is inflexible, and ends up being overly complicated. \par

\medskip 

The mathematical models of the Bell-LaPadula and Biba trusted system \cite{key-2,key-3} are complete and consistent. Typically, the issues which have arisen have arisen during the implementation of the models.
The implementation problems have not been the sets of semantics and mathematical rules to enable trust, but rather the underlying data
structures behind the rules are bit-vectors to represent categorical data. \par

\medskip 

The lack of uptake and efficacy of the Trusted System control over the years is primarily due to the inflexibility and maintenance complexity of using a simplistic use of bit-vectors when implementing the trusted systems model. \par

\medskip 

We introduce the concept of security tokens that encode various security attributes as vectors of integers that can be aggregated by summation or multiplication. These aggregated security tokens possess mathematical properties that maintain orthogonality across encoded attribute elements. Each element in the vector represents categorical security data, which can be easily decoded when required. The test of dominance between security tokens is facilitated by a simple aggregate operation, allowing high-performance filtering or switching without the need to decode the vectors. These tokens, while individually maintaining orthogonality across the encoded attributes that constitute a token, are stored and manipulated in their aggregated integer form, enabling efficient identification of overlaps between the aggregate security tokens. If overlaps are NOT detected between a user security token, who is requesting information, and the information security token, i.e., a zero subset exits between User and Information, then access is denied. \par

\medskip 

We have named this method Integer-Based Access Control (IBAC). This approach addresses the flexibility and complexity issues of earlier trusted systems and overcomes all the limitations of RBAC, ABAC, and CapBAC discussed earlier in the paper. In essence, IBAC surpasses the implementation limitations of bit vectors by utilising a superior bit vector abstraction while maintaining the performance advantages of raw bit vectors. When used in a trusted system, with some mathematical modifications to the model, trusted systems and MLS can overcome past limitations and provide a powerful tool for user authorisation to data and information. \par

\medskip 

Trusted systems could then be used as a highly performant, robust, and ubiquitous preventative authentication control, enabling trust on zero-trusted data sources by providing appropriate separation of duties based on user role separation when accessing data. \par

\medskip 

The area of interest and application of IBAC technology in this paper, i.e., the use case, includes the Relational Database Management System, and the data filtering for various Natural Language Processing (NLP) and AI retrieval methods. I.e., a focus on data protection for when using methods such as inverted lists and vector embeddings. \par

\medskip 

When we developed IBAC to resolve the authorisation data access control problem, a simple premise was devised: individuals within an organisation without authorisation to access certain portions of a 'data set' within a data source, whether it be a relational database, NoSQL database, vector database, file system, or even a message in a message queue, should have the unauthorised rows, documents, or messages filtered out before any data are returned in response to a query. Furthermore, the enforcement of access policies should be governed by a unified set of policy rules applicable to all data holdings in the organisation, ensuring consistency in authorisation and access control.\par

\medskip 

For any organisation only one single policy, i.e., set of trusted
system access rules, should drive the access or denial of 'information' requested by each 'user', across all of the various sources of information held by that organisation. \par

\medskip 

A set of corporate policy rules and supporting technology should be provided to allow access authorisation at the lowest level of granularity across any application, database, messaging layer, and network. All authorisation access to the data should be governed by a single policy across the entire technology stack. \par

\medskip 

Historically, after a user gains authenticated access to an application, we rely on the application functionality to ensure that access is appropriate in terms of separation of duty and need-to-know. The big idea and paradigm shift in this paper are two-fold. \par

\medskip 

Firstly, in addition to verification through application code testing, the IBAC model is mathematically sound, in contrast to ABAC and CapBAC. The completeness and consistency of the mathematical model ensure that as the organisation creates the data security policy, it is encoded in the technology, maintaining complete and consistent mathematical relations throughout the technology stack. This means that the policy is devised with confidence, free of security holes, and created in alignment with the organisational policy. Any change is an extension, an addition, rather than requiring extensive maintenance and code changes. \par

\medskip 

Secondly, IBAC shifts the responsibility for authorisation and data access from the application layer to the database layer. Traditionally, when access is managed at the application layer, it is merely hoped that application developers get the coding process right through extensive testing. In contrast to Application Code, IBAC provides controlled access through a central policy that is easily distributed across the technology stack. This approach ensures that a single policy is embedded and implemented within the fabric of each specific technology in the stack.

\medskip 

The authorised user's access should be established on a need-to-know
corporate policy basis, and then encoded into the data itself. Plus,
the implementation should be a simple control through a generic function
that would allow or deny access based on the users 'need-to-know',
enforcing clear separation of duty of users. \par 

\medskip 

\subsection{IBAC principles}

As stated earlier in developing IBAC, the traditional 'rules-based' direct
semantic implementation (for ABAC) using bit-vectors was abandoned because of
the inherent inflexibility, complexity, and lack of performance due
to problems with sparse arrays of the bit-vectors. An abstraction
of the bit vectors was instead developed, replacing the semantic bit vector
with orthogonal integer encoded tokens, plus a generic dominance relation
that exploited the orthogonality properties between the user (Subject)
and information (Object) to deny or allow access requests.  
The approach developed, \textquotedblleft orthogonal encoding\textquotedblright{}
and the mechanics are described in this paper, and it is an abstraction of the bit vector and is empirically superior for implementing the
trusted system.

\medskip 

As a metaphor if the reader is familiar with the concept of arithmetic coding for
lossless compression, then ``Orthogonal Encoding'' is a good parallel.

\section{Classical Implementation of the Trusted System}

Starting with the classical model of the "trusted system" and the
definition of the dominance relation, the paper uses the Amoroso \cite{key-4}
style of discrete mathematical description as the basis for the security model description for the trusted system semantics. 
\medskip 
A $Subject$ is defined as a request by some user of user role to
gain access to system information or resources 
\medskip 
An $Object$ is defined as a system repository where information or
resources are stored made available for access by the $Subject$ 
\medskip 
Security labels are the primary meta-data semantics associated with
either a $Subject$ or $Object$.
\begin{itemize}
\item Users as $Subjects$ are holders of a clearance token, i.e. a set
of semantic labels, to gain access to $Objects$ in terms of the
need-to-know and the need-to-share. 
\item The information $Objects$ and the resource are held within a system repository and are pieces of information that are tagged with metadata, a set of semantics, a security label, to control $Subject's$ access to the information. 
\item $Object$ labels (implemented as security tokens) can be viewed as
the policy that filters the User $object's$ request (also a security
token) by a mathematical relation called a $dominance$ relation. 
\item If the $dominance$ relation holds on a request, a user $Subject$
gains access to the information ($Object$ resource), else if the
dominance relation is false, access to the resources is denied to the
user.
\end{itemize}

\subsection{Security Dominance Relation definition}

What separates IBAC and trusted systems from ABAC and CapBAC is the mathematical model and in particular the mathematics of dominance. \par

\medskip

The dominance function is a fine-level granular filtering function that works by
checking to see if the user has a directed path to the data object that is being protected and requested by the user. In graph theory, dominance refers to the relationship between nodes (vertices) in a directed graph (digraph). \par

\medskip

Dominance is defined from graph theory: node $ a $ dominates node $ b $ if every control flow path from the entry node of the graph to node $ b $ must pass through node $ a $

\medskip

\FloatBarrier 
\begin{figure}[ht]
    \centering
    \includegraphics[scale=0.3]{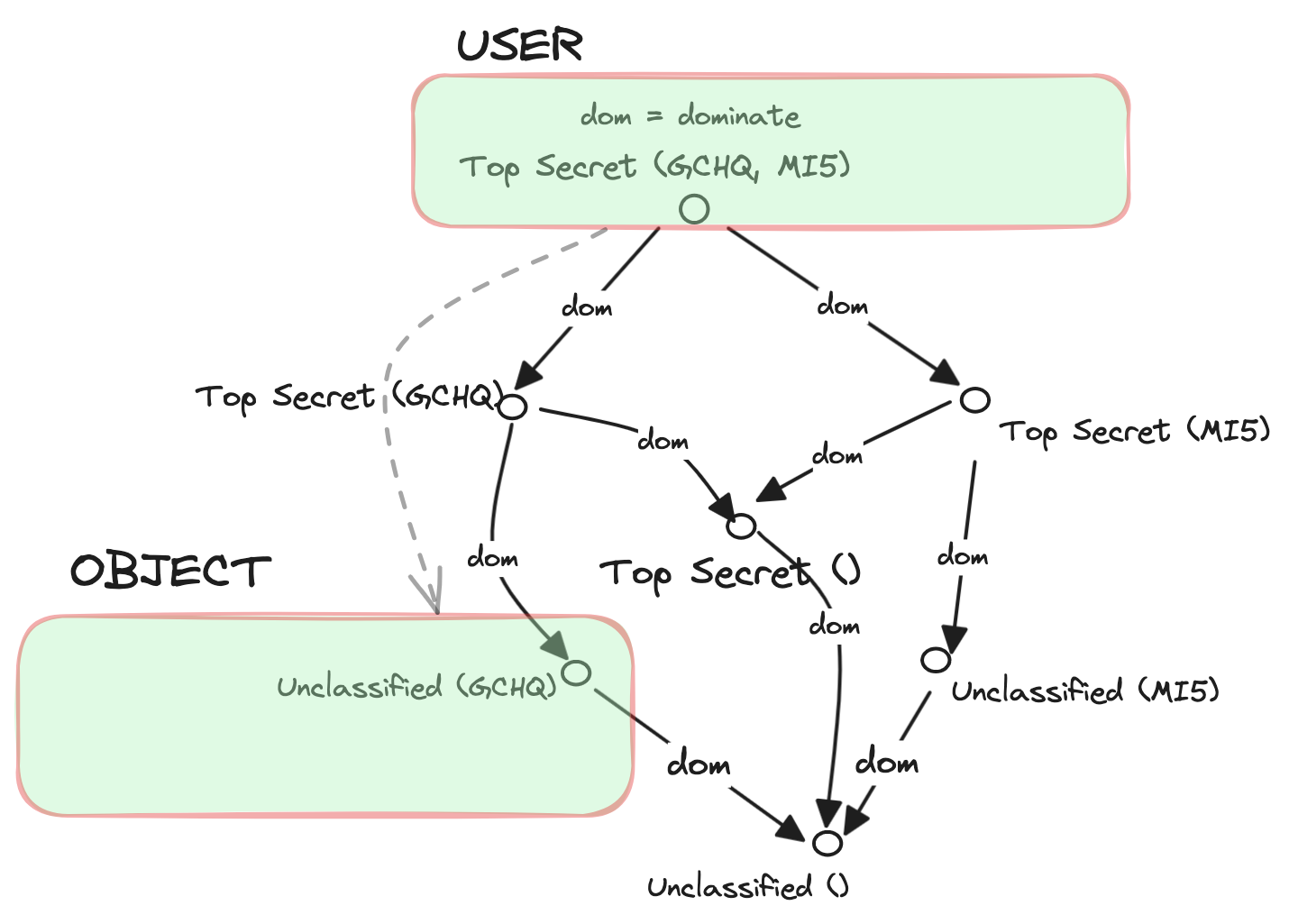} \par
    \textbf{diagram-1} graph dominance is True
\end{figure}

\FloatBarrier 

\medskip 

The above is an example of testing for dominance (dominance relation), that is, testing that there is a directed path between $ Subject $, the user, and $ Object $, the data being requested. In the above case, such a path exists, so the dominance relation between that user and object is True. \par

\medskip 

If the path cannot be traversed, we say that there is no dominance relation between the user and the object. Therefore, any request from the user is denied. \par

\medskip

\FloatBarrier 
\begin{figure}[ht]
    \centering
    \includegraphics[scale=0.3]{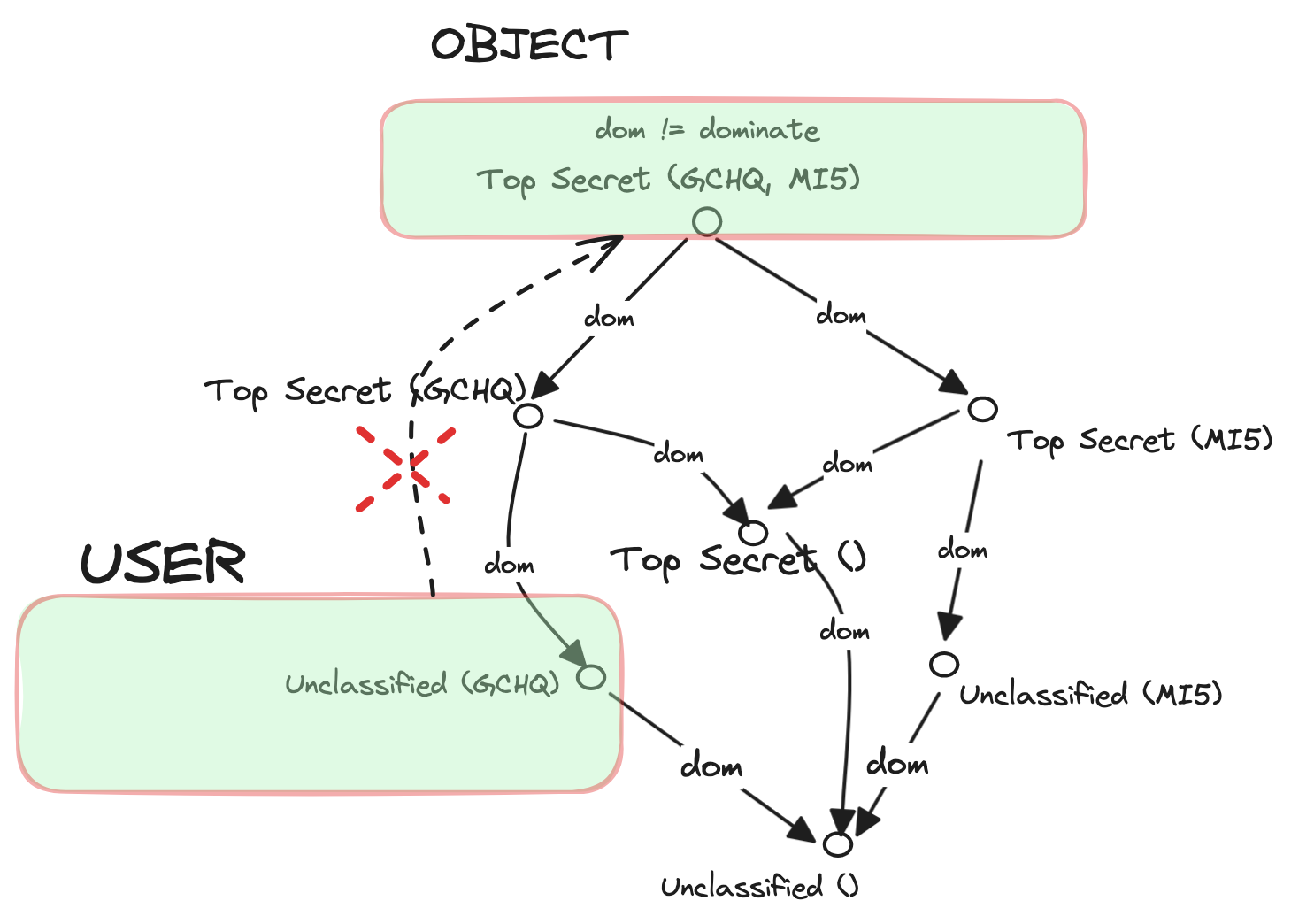} \par
    \textbf{diagram-2} graph dominance is False
\end{figure}

\FloatBarrier 

In the second case above, the dominance test fails because there is no directed graph (no path) between the user and the object. In the example the user cannot traverse the graph to the Object and so access is denied to that User for that object. in this case the dominance relation between that user and object is False. \par

\medskip

In summary Dominance is a graph concept which asks the question, can the user requesting access get to the label protecting data through a directed path. \par

\medskip

\subsection{Flattening the dominance structure - Proof of Equivalence Between Graph Dominance and Set Inclusion using a set encoding method }

All dominance testing methods are based on a flat dominance structure. Let us look at the first step of flattening the dominance structure by proving that hierarchical dominance in the directed graph is equivalent to a flattened structure, i.e., the subset relationships using a set encoding structure to provide set inclusion. \par

\subsubsection{Directed Graph Dominance}
Let $G = (V, E)$ be a directed graph with vertices $V$ and edges $E$.
For nodes $u, v \in V$, $u \rightarrow v$ indicates that $u$ dominates $v$.

\textbf{Dominance:} Node $u$ dominates node $v$ if every path from the start node $s$ to $v$ must go through $u$.

\subsubsection{Set Encoding}
Each node $u \in V$ is encoded as a set $S(u) \subseteq V$, where $S(u)$ contains $u$ and all nodes dominated by $u$.
The sets follow the subset relationship: $S(u) \supseteq S(v)$ if and only if $u \rightarrow v$.

\subsubsection{Proof of Equivalence}
We need to show that for any nodes $u, v \in V$:
\[
u \rightarrow v \text{ in the graph } G \text{ if and only if } S(u) \supseteq S(v) \text{ in the set encoding}.
\]

\subsubsection{1. Graph Dominance Implies Subset Relationship}
\textbf{Assumption:} $u \rightarrow v$

\textbf{To Prove:} $S(u) \supseteq S(v)$

By definition, $u \rightarrow v$ means every path from the start node $s$ to $v$ includes $u$.
Therefore, $v$ and all nodes dominated by $v$ are included in the set $S(u)$.
Hence, $S(u)$ includes $S(v)$, which means $S(u) \supseteq S(v)$.

\subsubsection{2. Subset Relationship Implies Graph Dominance}
\textbf{Assumption:} $S(u) \supseteq S(v)$

\textbf{To Prove:} $u \rightarrow v$

By definition, $S(u) \supseteq S(v)$ means that every node in $S(v)$ is also in $S(u)$.
This implies that all nodes dominated by $v$ are also dominated by $u$.
Therefore, $u$ must dominate $v$ in the graph $G$, i.e., $u \rightarrow v$.

\subsubsection{Application to Given Hierarchical Labels}

Given the nodes topSecret, Secret, and Public:
\[
\text{topSecret} \rightarrow \text{Secret}
\]
\[
\text{Secret} \rightarrow \text{Public}
\]

Set encoding:
\[
S(\text{topSecret}) = \{\text{topSecret}, \text{Secret}, \text{Public}\}
\]
\[
S(\text{Secret}) = \{\text{Secret}, \text{Public}\}
\]
\[
S(\text{Public}) = \{\text{Public}\}
\]

We show:
\[
\text{topSecret} \rightarrow \text{Secret} \implies S(\text{topSecret}) \supseteq S(\text{Secret})
\]
\[
S(\text{topSecret}) = \{\text{topSecret}, \text{Secret}, \text{Public}\}
\]
\[
S(\text{Secret}) = \{\text{Secret}, \text{Public}\}
\]
\[
\{\text{topSecret}, \text{Secret}, \text{Public}\} \supseteq \{\text{Secret}, \text{Public}\}
\]

\[
\text{Secret} \rightarrow \text{Public} \implies S(\text{Secret}) \supseteq S(\text{Public})
\]
\[
S(\text{Secret}) = \{\text{Secret}, \text{Public}\}
\]
\[
S(\text{Public}) = \{\text{Public}\}
\]
\[
\{\text{Secret}, \text{Public}\} \supseteq \{\text{Public}\}
\]

Thus, the hierarchical dominance in the graph is equivalently represented by the superset relationships in the set encoding.

\subsubsection{Conclusion}
We have shown that for any nodes $u$ and $v$:
\begin{itemize}
  \item If $u \rightarrow v$ in the graph, then $S(u) \supseteq S(v)$ in the set encoding.
  \item If $S(u) \supseteq S(v)$ in the set encoding, then $u \rightarrow v$ in the graph.
\end{itemize}

Therefore, the hierarchical dominance in the directed graph is equivalent to the superset relationships in the set encoding.

\subsection{Trusted System Security Policy - Mathematical Schema label definition }

Now that dominance has been defined, and we show how it is flattered, we provided an articulated example that follows a familiar scenario of metadata security labels and how these labels are used to provide access by a trusted system. We introduce the following label categories; security levels known as $levels$ and security agencies known as $ categories $, or as $ compartments $. Both are used to articulate the security semantics of the security model, which forms the organisational security policy. To help the reader,
the following terms can be interchanged:
\begin{itemize}
\item $Categories$ and $Compartments$.
\item $Subject$ and $User$
\item $Object$ and $Information$
\item $Semantics$ and $Labels$
\end{itemize}
A Security label is defined as follows: 
\medskip 

$labels=levels\times\wp(categories)$ 
\medskip 

That is, the set of labels is equal to a cross product of the set of
levels\textquoteright{} and the power set of 'categories\textquoteright  
\medskip 

The 'level' is a member of a set of classifications like \textquoteleft TopSecret\textquoteright ,
\textquoteleft Secret\textquoteright{} or \textquoteleft Protected\textquoteright{}
such that ordered pairs exist, that make up the graph. 
\medskip 

$TopSecret\geq Secret\geq Protected\geq Public$ 

 \medskip 

The category is a member of a power set of need-to-know / need-to-share
'compartments' like \textquoteleft GCHQ\textquoteright , \textquoteleft MI5\textquoteright ,
\textquoteleft MI6\textquoteright . 

 \medskip 

$\{GCHQ,MI5,MI6\}$ 

 \medskip 

An example of one particular security schema, known as a security
policy might be : 

 \medskip 

$levels=\{TopSecret,Secret,Protected,Public\}$ \par
$compartments=\{GCHQ,MI5,MI6\}$ \par
$\text{\ensuremath{\wp}}(compartments)=\{\text{\ensuremath{\varnothing}},\{GCHQ\},\{MI5\},\{GCHQ,MI5\}........etc\}$

 \medskip 

Continuing the example; three of the possible security label instances
from $levels\times\wp(compartments)$ that create the security policy
include: 

 \medskip 

$(TopSecret,\{GCHQ,MI5\})$ 
$(Secret,\{GCHQ,MI5,MI6\})$ 
$(Public,\{GCHQ\})$

\medskip

A binary relation on the set of ordered pair labels called dominates
is then introduced. I.e. a subset of the cross product of $ labels \times{} labels $: \par

\medskip

\noindent $ Subject\subseteq labels\wedge Object\subseteq labels $

\medskip

When an ordered pair of labels $ (Subject, Object)  $ is an element of
the dominant set of labels, we say that a dominance relation exists.

\medskip

$ (Subject,Object)\in dominate $

\medskip

When an ordered pair of labels $ (Subject, Object) $ is not an element
of the set of labels that dominates, we say that a dominance relation
does not exist. 

\medskip

$ (Subject,Object)\notin dominate$

\medskip

The dominance relation itself is defined as follows: \par

\medskip

$\forall (x_1 \in \text{Subjects}, \ x_2 \in \text{Objects}):(x_1, x_2) \in \text{dominates} \iff$ \par
$\text{levels}(x1) \geq \text{levels}(x2) \ \wedge \
\text{compartments}(x1) \supseteq \text{compartments}(x2) $ \par

\medskip 

The dominance relation that is true is a binary relation $(a,b)$
on the set of labels such that when the relation is maintained, the level
of $a$ is greater than (or equal) to the level of $b$ and the compartments
of $a$ are a superset of the compartments of $b$ 

 \medskip 

Hence, the following are all statements of the dominance binary relation
from the security schema previously defined: \par

\medskip

$ ((Secret,\{GCHQ,MI5,MI6\}),(Secret,\{MI6\}))\in
dominates \\ \\
((TopSecret,\{MI6\}),(Secret,\{MI5\}))\notin dominates $
 \medskip 

Note: the second example above is false, i.e. $((TopSecret,\{MI6\}),\ (Secret,\{MI5\}))\notin dominates$
as the object's compartment $MI5\notin$ in $(TopSecret,\{MI6\})$ of the subject.

\subsection{Simplification of the Trusted System security policy by set Inclusion and a simple test for subset between $ user $ and $ object$ }

There are many performance challenges when implementing a classical
model of a ``dominance relation'' using a set of semantic rules. 

 \medskip 

The first step is the classical semantic model itself can be easily
mathematically reduced by set inclusion.  

 \medskip 

As we have proven by the carefully encoding of the security labels; the dominance relation can be reduced to a pure subset. 

 \medskip 

A finite set of A labels is ``included'' in the set B of labels
(a subset of B), $A\subseteq B$, if and only if the cardinality of
the A and B intersection is equal to the cardinality of A\cite{key-5}.

\medskip
$A\subseteq B\:\:iff\:\:(A\cap B=A)$
\medskip

In addition, for any set of security labels S, the property of the
inclusion relation in the set is a partial order of $\wp(S)$, that is,
a partial order of the power set of $ S $.

\medskip
$A\leq B\iff A\subseteq B$ \par
$(a,\text{b})=\{\{a\},\{\text{a},\text{b\}\}}$ 
\medskip

Extending the example above, the set of levels $levels=TopSecret\geq Secret\geq Protected\geq Public$
is in fact an inclusion:

\medskip
$ \{ \text{TopSecret}, \text{Secret}, \text{Protected}, \text{Public} \} $ \par
$ \supseteq \{ \text{Secret}, \text{Protected}, \text{Public} \} $ \par
$ \supseteq \{ \text{Protected}, \text{Public} \} $ \par
$ \supseteq \{ \text{Public} \} $ \par
\medskip

$\{\{TopSecret,Secret,Protected,Public\}\supseteq\{Secret,Protected,Public\}\supseteq\{Protected,Public\}\supseteq\{Public\}\}$
\medskip

The complete security scheme then showing the inclusion of the security
levels joined with the compartments becomes: \par

\medskip
$\{\{TopSecret,Secret,Protected,Public\}\supseteq\ \{Secret,Protected,Public\}\supseteq\{Protected,Public\}\supseteq\{Public\}\} $ \par
$\cup \ \{GCHQ,MI5,MI6\}$ \par
\medskip 

By careful semantic encoding, the dominance relation can then be reduced
to just testing for subsets of levels and subsets of compartments: \par

\medskip
$ {\forall}(x1\in Subjects,x2\in Objects):(x1,x2)\in dominates\iff $ \par
$ (levels(x1)\supseteq levels(x2)\:\wedge 
compartments(x1)
\supseteq compartments(x2) $\par
\medskip

The implication for the efficient implementation of the dominance
relation is then a simple test for subsets. \par

\medskip
$ (SecLabels=levels\cup compartments)\:\wedge$ \par

$ {\forall}(x1\in Subjects,x2\in Objects):(x1,x2)\in dominates\iff $
$ (SecLabels(x1)\supseteq SecLabels(x2) $ \par
\medskip

The reduction of the dominance relation to a simple test for subsets after carefully encoding the security labels provides massive performance gain when implementing the dominance relation using aggregate functions, which will be discussed later in the paper. \par
\medskip

\section{IBAC Process Dominance - Extending the trusted system's model}
We made the point earlier that moving away from the mathematically sound trusted Bell-LaPadula and Biba models to build access control systems such as ABAC and CapBAC, which lack a solid mathematical foundation, was a mistake. Instead, extending the Bell-LaPadula and Biba mathematics to enhance robustness and introduce flexibility should have been the focus. \par

\medskip

From now on, when we refer to Bell-LaPadula, we refer to Bell-LaPadula including the Biba model as well. \par

\medskip

We outline an approach here to extend Bell-LaPadula based on some mathematical research that was developed in the 1990s (see Hargraves). \par

\medskip

Hargraves makes the point in his master's dissertation that Bell-LaPadula can be made more flexible if both the Subject dominates the Object and the Process dominates the Object. \par

\medskip

This is an interesting philosophical shift. The argument being a Subject can only gain access to an Object by some Process, so why is the Process itself not part of the dominance relation and not part of the Bell-LaPadula trusted system model. \par

\medskip 

Hargraves \cite{key-12} indicates that both the Subject and the Process should dominate the object \par

\medskip

We will show that by adding the additional process constraint to the Bell-LaPadula model, the model becomes much more flexible and addresses the original disadvantages of Bell-LaPadula that led to the development of ABAC and CapBAC.  \par


\medskip
We describe the extended Bell-LaPadula model and the associated proof using category theory \cite{key-7}. Category theory is an abstraction of the predicate logic that enables us to express more efficiently and succinctly the necessary mathematics.  \par   

\medskip

The key point to the Bell-LaPadula extension is that we introduce one-to-one (bijective) relationship mappings between the "Subject" and a "Process" which enables greater flexibility in the model when accessing "objects". \par

\medskip

So in addition to user as the Subject mapped to the Object, we also have combinations of Subjects and Processes, as tuples mapped to the object. \par

\medskip
   
For example, $("user1", "printer", "secret") \ and \ ("user2", "briefing\_room", "top secret") $. \par
 
\medskip

So in addition to a Subject in this case, the User dominating the Object, so does the (User, Process ) tuple dominate the object. \par

\medskip

For example, $ g: \text{("user1", "printer", "top secret")} \to \text{("object1", "secret")} \text{ where } g \in \text{dominates} $. \par

\medskip
and
\medskip

$ f: \text{("user2", "printer", "secret")} \to \text{("object2", "top secret")} \text{ where } f \notin \text{dominates} $ \par
\medskip

A more formal definition with proof of the extended Bell-LaPadula is as follows.

\subsection{One-to-One  "Subject" and a "Process" (Bijective) Relationship }

The key to the Bell-LaPadula extension is to introduce a bijection between Subject ($S$) and Process ($P$).  In category theory, the bijective relationship can be represented with a commutative diagram showing that a morphism \(f: S \to P\) has an inverse \(g: P \to S\) such that \(g \circ f = \text{id}_s\) and \(f \circ g = \text{id}_p\):

\[
\begin{tikzcd}
S \arrow[r, shift left, "f"] & P \arrow[l, shift left, "g"] \\
S \arrow[u, "\text{id}_S", rightarrow] & P \arrow[u, "\text{id}_P", rightarrow]
\end{tikzcd}
\]

This diagram indicates that \(g \circ f = \text{id}_S\) and \(f \circ g = \text{id}_P\), demonstrating that \(f\) is bijective. \par

\medskip

This can then be used to extend the use cases where Bell-LaPadula with the additional Process constraint provide greater flexibility in terms of how the model can be used for access control. \par

\medskip

From an implementation perspective of the bijective relationships, the User/Process mappings are setup as tuples (user, process) by the organisation, mappings prior to any usage .

\subsection{Formal definition Bell-LaPadula Extension with example problem-1 of a secret printer printing lower classification documents }

\subsubsection {Problem Statement}
\begin{itemize}
  \item \textbf{Process \( P \)}: 'Printer' with classification 'Top Secret'
  \item \textbf{Subject \( U \)}: 'User' with 'Top Secret' clearance wants to print a 'document'.
  \item \textbf{Object \( O \)}: 'Document' being requested for printing is 'Public'
\end{itemize}

According to the Bell-LaPadula/Biba model, a 'Top Secret' process like a printer should not write to print a 'Public' object like a document. \par

\medskip

From an infrastructure perspective, this is a financial waste to double or even triple infrastructure costs. \par

\medskip

The problem can be overcome by extending Bell-LaPadula/Biba to include an additional process dominance relation, in this case the printer being the process that also dominates the object. \par

\medskip

\subsubsection{Solution}
\begin{itemize}
  \item There exists a 1-to-1 mapping between the user \( u \) and the process \( p \),  \\
  denoted as \((U, P, \text{"User/Process classification"})\). These 1-to-1 mappings are setup prior to when access is required by the user and setup by an independent 3rd party administrator based on appropriate change control for that organisation. 
  \item This mapping allows the user \( U \) to use the printer \( P \) for writing (printing a document), assuming that the user \( U \) has an appropriate classification and dominates the object.
  \item Commutativity:
    \begin{align*}
      U \rightarrow P & \quad \text{(User has the allowed access to the Printer, set up prior as a bijection tuple)} \\
      P \rightarrow O & \quad \text{(the user/printer bijection tuple has allowed access to write to the Object)} \\
      U \rightarrow O & \quad \text{(The user  has direct read access, i.e., they dominate the document)}
    \end{align*}
\end{itemize}

\subsubsection{Formalization in Category Theory}

\subsubsection*{Definitions and Notations}
\textbf{Objects}:
\begin{itemize}
  \item \( UP = \{(U, P, \text{"User/Process classification"}) \} \) (Set of users in context and their associated processes)
  \item \( U = \{(P, \text{"User classification"}) \} \) (Set of users and their classifications)
  \item \( O = \{(O, \text{"Object classification"}) \} \) (Set of objects and their classifications)
\end{itemize}

\textbf{Morphisms}:
\begin{itemize}
  \item \( y: U \to UP \) (Mapping from Users to Users in context of Processes, bijective)
  \item \( y^{-1}: UP \to U \) (Inverse mapping from Users in context of Processes to Users)
  \item \( f: UP \to O \) (Mapping from Users in context of Processes to Objects)
  \item \( h: U \to O \) (Composite mapping from Users to Objects, defined as \( h = f \circ y \), for allowed write access via the printer)
  \item \( d: U \to O \) (Direct mapping from Users to Objects, assuming allowed User read access)
\end{itemize}

\subsubsection*{Compositionality and Commutativity}
\begin{itemize}
  \item The composite function \( h \) is defined as \( h = f \circ y \).
  \item Commutativity ensures that:
    \[
    h(U) = f(y(U))
    \]
    This means:
    \[
    h(U, \text{"User classification"}) dom (O, \text{"Object classification"})
    \]
    if:
    \[
    f((U, P, \text{"User/Process classification"})) dom (O, \text{"Object classification"})
    \]
    and:
    \[
    d((U, \text{"User classification"})) dom (O, \text{"Object classification"})
    \]
  \item Conditional access: The user \( U \) can access the object \( O \) assuming direct read access via \( d \) allowing indirectly write access via \( H \) using a secret printer for lower-classified documents.
\end{itemize}

\subsubsection*{Commutative Diagram}
\[
\begin{tikzcd}
U \arrow[r, "y"] \arrow[dr, "h"'] \arrow[rr, "d", bend left] & UP \arrow[r, "f"] \arrow[d, "y^{-1}"] & O \\
& U \arrow[ur, "h"']
\end{tikzcd}
\]

This diagram represents the commutative properties of the mappings, ensuring that \( U \rightarrow P \) and \( P \rightarrow O \) commute properly to allow \( U \rightarrow O \). \par

\medskip

Thus, the user \( U \) writes the object \( O \) assuming they have (1) direct read access dominance and (2) through the combination (user / printer) write access process through \( P \), depending on the conditions set by the access control policies of the tuples specified of the bijective mappings of the user and process combinations, for example $(user1, \ printer1)$.

\subsection{Formal Definition Bell-LaPadula Extension with Example Problem-2 of a Briefing Room}

\subsubsection{Problem Statement}
\begin{itemize}
  \item \textbf{Process \( R \)}: Briefing room to brief low-ranking officers
  \item \textbf{High-ranking officer \( H \)}: Holds a plan \( P \) at 'Top Secret'
  \item \textbf{Low-ranking officer \( L \)}: Can't see the plan \( P \) as his ranking is only 'Secret'
\end{itemize}
In the briefing room \( R \), the high-ranking officer \( H \) who holds the plan \( P \) can disclose the plan \( P \) to the low-ranking officer \( L \) while in the room \( R \).

\subsubsection{Solution}
\begin{itemize}
  \item \textbf{Tuples}: \((H, R)\) and \((L, R)\), both are allowed in the room for briefings. setup prior by an Administrator.
  \item \textbf{Mappings}:
    \begin{itemize}
      \item \( k: H \to P \) (High-ranking officer \( H \) has prior access to the plan \( P \), i.e., dominates)
      \item \( f: (H, R) \to P \) (High-ranking officer \( H \) can disclose the plan in the room \( R \))
      \item \( h: (H, R) \to (L, R) \) (High-ranking officer \( H \) in the room \( R \) can disclose to low-ranking officer \( L \) in the room \( R \))
      \item Therefore, \( j: (L, R) \to P \) (Low-ranking officer \( L \) can see the plan in the room \( R \))
    \end{itemize}
\end{itemize}

\subsubsection{Formalization in Category Theory}

\subsubsection*{Definitions and Notations}
\textbf{Objects}:
\begin{itemize}
  \item \( HR = \{(H, R, \text{"High-ranking officer/Room classification"}) \} \) (Set of high-ranking officers in context and their associated rooms)
  \item \( LR = \{(L, R, \text{"Low-ranking officer/Room classification"}) \} \) (Set of low-ranking officers in context and their associated rooms)
  \item \( H = \{(H, \text{"High-ranking officer classification"}) \} \) (Set of high-ranking officers and their classifications)
  \item \( L = \{(L, \text{"Low-ranking officer classification"}) \} \) (Set of low-ranking officers and their classifications)
  \item \( P = \{(P, \text{"Plan classification"}) \} \) (Set of plans and their classifications)
\end{itemize}

\textbf{Morphisms}:
\begin{itemize}
  \item \( y: H \to HR \) (Mapping from high-ranking officers to high-ranking officers in context of rooms, bijective)
  \item \( y^{-1}: LR \to H \) (Inverse mapping from low-ranking officers in context of rooms to high-ranking officers)
  \item \( f: HR \to P \) (Mapping from high-ranking officers in context of rooms to plans)
  \item \( h: H \to LR \) (Mapping from high-ranking officers to low-ranking officers in context of rooms)
  \item \( j: LR \to P \) (Mapping from low-ranking officers in context of rooms to plans)
  \item \( k: H \to P \) (Mapping from high-ranking officers who have prior access to the plan)
\end{itemize}

\textbf{Compositionality and Commutativity}
\begin{itemize}
  \item The composite function \( h \) is defined as \( j\circ h = f \circ y \).
  \item Commutativity ensures that:
    \[
    f(y(H, R) = j(h(L, R))
    \]
    This means:
    \[
    h(L, R, \text{"low-ranking officer/Room classification"}) dom (P, \text{"Plan classification"})
    \]
    if:
    \[
    f((H, R, \text{"High-ranking officer/Room classification"})) dom (P, \text{"Plan classification"})
    \]
    and:
    \[
    k((H \text{"High-ranking officer"})) dom (P, \text{"Plan classification"})
    \]
  \item Conditional access: The low-ranking officer \( L \) can see the plan \( P \) in the room \( R \), given the context and classification constraints.
\end{itemize}

\subsubsection{Commutative Diagram}
\[
\begin{tikzcd}
H \arrow[r, "y"] \arrow[dr, "h"'] \arrow[rr, "k", bend left] & HR \arrow[r, "f"] \arrow[d, "y^{-1}"] & P \\
& LR \arrow[ur, "j"']
\end{tikzcd}
\]

This diagram represents the commutative properties of the mappings, ensuring that \( H \rightarrow R \) and \( R \rightarrow P \) commute properly to allow \( L \rightarrow P \) in the room \( R \). \par

\medskip

Thus, the high-ranking officer \( H \) discloses the plan \( P \) to the low-ranking officer \( L \) within the briefing room \( R \), assuming the high-ranking officer \( H \)  has (1) direct classification dominance and (2) conditional access within the context of the room exists for both the high-ranking officer \( H \)  and the low-ranking officer \( L \).

\section{Implementation of dominance} 
\subsection{The problem of bit-vector implementation}

Bit-vectors would seem a natural fit for implementing the notion of
trust by substituting each security semantic for a bit position in
the bit-vector. In fact, the observation of inclusion that $(A\subseteq B)\:\:iff\:\:(A\cap B=A)$
can be directly and easily applied to build a dominance relation between two
bit-vectors A and B. This is the classical way of encoding and testing security tokens for dominance. 

 \medskip 

In this paper only two sets of bit-vectors of $(user,object)$ are used in this for the demonstration and testing of dominance between $(user,object)$.

 \medskip 

In terms of bit-vector testing for Dominance, the user only obtains access to information
if the relation $(A\subseteq B)\:\:iff\:\:(A\cap B=A)$ of the bit-vectors
hold 

 \medskip 

Their are two relations used in this paper time and again for demonstration
and being tested using different encoding techniques to ensure $(user,object)\in dominates$
are as follows:
\begin{itemize}
\item $((Secret,\{MI5,MI6\}),(Secret,\{MI5\}))\in dominates$  

 \medskip 

shown in its included form: $(\{Secret,Protected,Public\}\cup\{M15,MI6\}),(\{Secret\}\cup\{MI5\})\in dominates$ 
\item $((Secret,\{MI5,MI6\}),(Secret,\{GCHQ,MI6\}))\notin dominates$ 

 \medskip 

shown in its included form: $(\{Secret,Protected,Public\}\cup\{MI5,MI6\}),(\{Secret\}\cup\{GCHQ,MI5\})\notin dominates$
\end{itemize}
It is important to note the following in regard to the examples above: 
\begin{itemize}
\item Only the $Subject$ i.e. the $User$ needs to be in the 'included
form' as the $User's$ label are the markings in terms of a holder
of a clearance level that contains access to other clearance levels,
and so on, i.e., hierarchical levels for user label markings. 
\item On the other hand the $Object$, the marking on the $Information$
is tagged with only \textbf{a singular }level in terms of it's classification
marking $(TopSecret\lor Secret\lor Protected\lor Public)$. There
is no concept of hierarchical levels for information label markings.
\item The two dominance relations above will be used throughout the whole
paper, to help articulate the various dominance mechanics and options,
i.e.: 
\begin{itemize}
\item 'P\_rel' {[}$\in dominates${]}, i.e. $((Secret,\{MI5,MI6\}),(Secret,\{MI5\}))\in dominates$
and
\item 'N\_rel' {[}$\notin dominates${]}, i.e. $((Secret,\{MI5,MI6\}),(Secret,\{GCHQ,MI6\}))\notin dominates$
\end{itemize}
\end{itemize}

\subsubsection{Dominance binary relation implementation as bit-vector boolean operations}

The dominance relation is implemented for bit-vectors as a boolean
operator, and is a function that applies the conjunctive, i.e., the
'and' operator, and then post the application of the 'and' operator,
then a test, if the resultant bit-vector of $user\land object=object$,
then $user$ is granted access, else $user$ access is denied. \par

\medskip 

\textbf{P\_rel:} \par
\begin{tabular}{l|l|l|l|}
\hline 
\textbf{Access Granted} & \textbf{user} & \textbf{object1} & \textbf{user $\land$ object1 = object1} \\
\hline 
Top secret & 0 & 0 & 0  \\
\hline 
Secret & 1 & 1 & 1  \\
\hline 
Protected & 1 & 0 & 0  \\
\hline 
Public & 1 & 0 & 0  \\
\hline 
GCHQ & 0 & 0 & 0  \\
\hline 
MI5 & 1 & 1 & 1  \\
\hline 
MI6 & 1 & 0 & 0  \\
\hline 
\end{tabular}  \par
\textbf{Figure-1}: Boolean dominance operation of: \par
$((\text{Secret}, \{\text{MI5}, \text{MI6}\}), (\text{Secret}, \{\text{MI5}\}))
\in \text{dominates}$ \par
\medskip 
Note: $ \text{Column(object1)} = \text{Column(user} \land \text{object1 = object1)}$

\medskip 

\textbf{N\_rel:} \par
\begin{tabular}{l|l|l|l|}
\hline 
\textbf{Access Denied} & \textbf{user}  & \textbf{object2} & \textbf{user $\land$ object2 $\neq$ object2}\\
\hline 
Top secret & 0 & 0 & 0 \\
\hline 
Secret & 1 & 1 & 1 \\
\hline 
Protected  & 1 & 0 & 0 \\
\hline 
Public & 1 & 0 & 0 \\
\hline 
GCHQ  & 0 & 1 & 0 \\
\hline 
MI5 & 1 & 0 & 0 \\
\hline 
MI6 & 1 & 1 & 1 \\
\hline 
\end{tabular} \par
\textbf{figure-2}: boolean dominance operation of: \par 
$((\text{Secret}, \{\text{M15}, \text{MI6}\}), (\text{Secret},
\{\text{GCHQ}, \text{MI6}\})) 
\notin \text{dominates}$ \par
\medskip 
Note1: $Column(object2)\neq Column(user\land object2\neq object2)$ \par
\medskip 
Note2: The resultant vector does not signal in itself the relation
$\neg dominate$, It is a secondary post binary operation of comparison
that signals that $(User$$,Object)\notin dominates$  \par

 \medskip 

bit-vectors have been the typical encoding scheme for implementing
trusted systems for the last 50 years. bit-vectors represent the simplest
of encoding scheme. On face value, they appear as a pragmatic and
straightforward encoding scheme that one could ever envisage. Each
bit indicates a security label semantic being set either to ``off''
or ``on''. It would appear that bit-vectors are just a simple representation
of a set of categorical data. 

 \medskip 

The trusted system semantics is certainly categorical; however, the
need for a highly performant implementation such as filtering billions
of rows in a database, and the continual maintenance of billions of
rows in a database, might not make the bit-vector a great choice. 

\subsubsection{bit-vectors are easy to hack}

The first problem with bit-vectors is that they are too easy to hack.
A bit patten of all 1's for the $User$ will dominate any $Object$.
Hence the problem of circumventing dominance when using bit-vectors
is very real. Any Mersenne number $(2^{p}-1)$ will be a bit pattern
of all 1's that will dominate any other bit pattern of the same or
lesser array size. See section 4 above, $A\subseteq B\:\:iff\:\:((A\cap B=A)$. 

 \medskip 

Object 1 above is dominated by the Mersenne number $(2^{7}-1)$, i.e.
$[1,1,1,1,1,1,1]$ 
$[TopSecret,Secret,Protected,Public,GCHQ,MI5,MI6]\iff[1,1,1,1,1,1,1]$

 \medskip 

$[TopSecret,Secret,Public,MI5,MI6]\iff[0,1,0,0,1,0,1]$ 

 \medskip 

$[1,1,1,1,1,1,1]\land[0,1,0,0,1,0,1]=[0,1,0,0,1,0,1]$ 

 \medskip 

Strict controls of managing $Subject$ and $Object$ bit-vectors are
therefore necessary for any implementation to ensure the user role
profiles and information markings are never tampered with or made
public. All $Subjects$ and $Objects$ (bit-vector) at rest without
encryption can be compromised.  

 \medskip 

Even more restrictive is the bit-vectors can never be transmitted
without encryption being employed as an additional layer of control.
Leading to performance overheads that will be experienced immediately
when testing for dominance. 

 \medskip 

The reader will see later with the Orthogonal Encoding methods; The
hacking problem disappears to the extent that Orthogonal Encoding
is more \textbf{secure} when storing and transmitting security tokens. 

\subsubsection{bit-vectors are difficult to manage and maintain and there is a sparse
array problem to overcome }

Typically in any large organisation, there may be 100+ bits or more
to manage, i.e., the bit\\ representations of the various security semantics.
As the enterprise security policy gets more complex, more and more
bits are then required, and they need to be managed.  

 \medskip 

The second problem is you need to manage both the 1s and 0s in the
bit-vector. You end up managing the inefficiencies of a sparse array
of bits.  

 \medskip 

The reader will see later with the Orthogonal Encoding methods, that
the sparse array problem disappears.

\subsubsection{What happens when you want to exceed the size of the bit-vector}

Closely aligned with 4.1.3, is a bit management issue. What happens
when the bit-vector of a particular size (vector length) that has
been implemented across all of your code, in all your security products,
now needs just one extra bit? You are forced to recode and then recompile
your entire code base. 

 \medskip 

The reader will see later with the orthogonal encoding methods that the
vector length problem also disappears.

\section{Bit-vector Orthogonal Subspaces of security vectors - the first step in fast
implementation methods }

\subsection {bits as unit vectors}
The security bit-vector can be considered from the perspective of
linear algebra. By definition, each possible bit combination of a
position of 1 with all other positions set to 0, can be interpreted
as a unit vector. Unit vectors providing the basis of a vector space
for the security 'domain of discourse'.  

 \medskip 

For example, assuming three security semantics of 'A', 'B' and 'C',
and then binary encoding of these three semantics $[A=001,B=010,C=100]$,
the binary encoding will result in three bit-vectors. The bit-vectors
can be viewed as three unit vectors $[[0,0,1],[0,1,0],[1,0,0]]$ ,
these unit vectors can then be considered the basis of a vector space
$\mathbf{R}^{\mathbf{3}}$ that stands for a universe of discourse
for a set of security semantics, in this case 'A', 'B' and 'C'. 

 \medskip 

By definition each of the three encodings $[[0,0,1],[0,1,0],[1,0,0]]$
are orthogonal to any other. A good question then is, comparing the
$user$ bit-vector with an $object$ bit-vector is it possible to
utilize the linear algebra convention of orthogonality to help in
the determination of the dominance between bit-vectors? 

 \medskip 

It turns out that by using the complement of the $user$ bit-vector, the
concept of orthogonality can be directly applied to test for dominance by the use of the dot product. 

\subsection{Dot Product as a test of Dominance}

Using the complement of the $user$ bit-vector, and the dot product operstion, a direct test of
orthogonality signifies dominance between $user$ and $object$.
That is, the dot product can be directly applied as a test for the dominance
relation, ${a\bullet b}=\sum_{i=1}^{n}a_{i}b_{i}=a_{1}b_{1}+a_{2}b_{2}+...\:a_{n}b_{n}$ 

 \medskip 

\textbf{P\_rel:} 
The rationale for P\_rel is as follows in the case where the dominance
relation is found between $user$ and $object$ : 

 \medskip 

If the relation of dominance holds for $(user,object1)\in dominate$,
then as $object1$ is a subset of $user$, there must be zero overlap
with the bit 'complement of the user' $(user^{C})$  

 \medskip 

The diagram below indicates that the "proper subset" using the vector dot product
function expects the result to be \textbf{zero} 

 \medskip 

i.e. ${user^{C}}\bullet{object1}=0$  \par
\includegraphics[width=3cm,height=3cm]{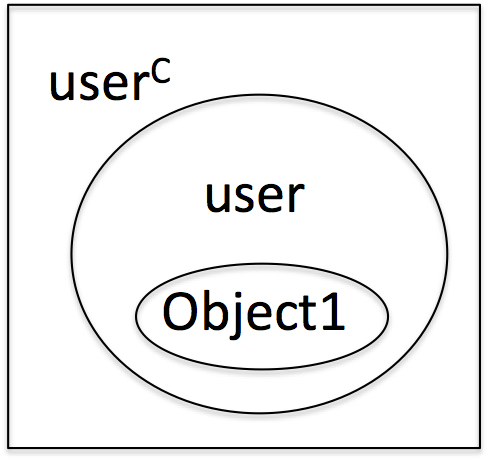} 

 \medskip 

\begin{tabular}{|l||l||l||l|l|}
\hline 
\textbf{Access Granted} & $user$ & $user^{C}$ & $object1$ & $user^{C}\bullet object1$\tabularnewline
\hline 
Top secret & 0 & 1 & 0 & 0\tabularnewline
\hline 
Secret & 1 & 0 & 1 & 0\tabularnewline
\hline 
Protected  & 1 & 0 & 0 & 0\tabularnewline
\hline 
Public & 1 & 0 & 0 & 0\tabularnewline
\hline 
GCHQ  & 0 & 1 & 0 & 0\tabularnewline
\hline 
MI5 & 1 & 0 & 1 & 0\tabularnewline
\hline 
MI6 & 1 & 0 & 0 & 0\tabularnewline
\hline 
\multicolumn{1}{|l|}{$\mathbf{\sum bits\Rightarrow=\perp}$} & \multicolumn{1}{l}{} & \multicolumn{1}{l}{} &  & \textbf{0}\tabularnewline
\hline 
\end{tabular} \par
\textbf{figure}-3: boolean dominance operation using dot product \\${user^{C}}\bullet{object1}=0$
: 
 $((Secret,\{MI5,MI6\}),(Secret,\{MI5\}))\in dominates$ 

 \medskip 

\textbf{N\_rel:} 
Conversely, the rationale for N\_rel is as follows in the case where
the dominance relation between the $user$ and $object$ does not
hold: 

 \medskip 

If the relation of dominance does not hold for $(user,object2)\notin dominate$,
then as $object2$ overlaps with the $user$ set, there will be an
encroachment (an overlap) with the 'complement of the user' $(user^{C})$

 \medskip 

 \medskip 

The diagram below indicates the encroachment in $user^{C}$ using
the vector dot product function expecting the result being \textbf{greater
than} zero. \par

 \medskip 

i.e. ${user^{C}}\bullet{object2}>0$ \par
\includegraphics[width=3cm,height=3cm]{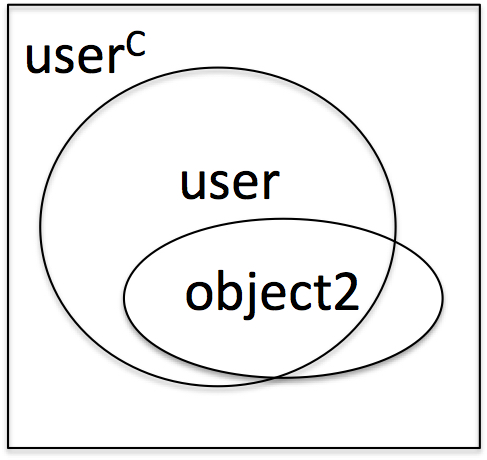} 

 \medskip 

\begin{tabular}{|l||l||l||l|l|}
\hline 
\textbf{Access Denied} & $user$ & $user^{C}$ & $object1$ & $user^{C}\bullet object2$\tabularnewline
\hline 
Top secret & 0 & 1 & 0 & 0\tabularnewline
\hline 
Secret & 1 & 0 & 1 & 0\tabularnewline
\hline 
Protected  & 1 & 0 & 0 & 0\tabularnewline
\hline 
Public & 1 & 0 & 0 & 0\tabularnewline
\hline 
GCHQ  & 0 & 1 & 1 & 1\tabularnewline
\hline 
MI5 & 1 & 0 & 0 & 0\tabularnewline
\hline 
MI6 & 1 & 0 & 1 & 0\tabularnewline
\hline 
\multicolumn{1}{|l|}{$\mathbf{\sum bits\Rightarrow\neq\perp}$} & \multicolumn{1}{l}{} & \multicolumn{1}{l}{} &  & \textbf{1}\tabularnewline
\hline 
\end{tabular} \par
\textbf{figure}-4: boolean dominance operation using dot product \\ $ {user^{C}}\bullet{object2}=1$
: 
 $((Secret,\{MI5,MI6\}),(Secret,\{GCHQ,MI6\}))\notin dominates$ 

 \medskip 

What the orthogonal discussion tells us is as follows, in addition
to looking at bit-vectors in a boolean logic sense, bit-vectors are
subject to the mathematical laws of linear algebra. This can be exploited
to test for dominance by using a simple dot product operation between the complement of $User$ and $Object$. If an encroachment exists in the set
$user^{c}$, i.e., $user^{C}\bullet object>1$, then by definition
the binary dominance relation does not hold.  

 \medskip 

There is a deeper level of connection between Boolean logic
and the dot product. It can be observed from the above two figures
3 \& 4 that $(a_{1}b_{1}+a_{2}b_{2}+...\:a_{n}b_{n}=a\bullet b)$,
where $a=user^{c}$ and $b=object$, such that there exists the
following conjecture, $({u_{x}.o_{x}\in user^{C}}\bullet{object)=}(u\in user^{C}\land o\in object)$.
The proof is as follows:

\subsubsection{proof: Dot Product as a Dominance Test - Equivalence between Logical Expressions for \( user \land object1 = object1 \) and \( user^C \bullet object1 = 0 \),  Plus \( user \land object2 \neq object2 \) and \( user^C \bullet object2 = 1 \) }

We aim to prove the first equivalence between the two logical expressions: \\ first between \( user \land object1 = object1 \) and \( user^C \bullet object1 = 0 \), and then   \\
second between  \( user \land object2 \neq object2 \) and \( user^C \bullet object1 = 1 \).

\subsubsection{Definitions and Initial Setup for $ user^C \bullet object1 = 0 $}

\begin{itemize}
    \item \( user^C \): The complement of user, representing all elements not in user.
    \item Dot Product: In this context, it is a boolean operation indicating the intersection. For boolean values, \( A \bullet B = 1 \) if both \( A \) and \( B \) are 1; otherwise, it is 0.
    \item Conjunction: \( \land \) represents the logical AND operation.
\end{itemize}

\subsubsection{Proof Steps for  - $ user^C \bullet object1 = 0 $}

We need to show that:
\[ user \land object1 = object1 \iff user^C \bullet object1 = 0 \]

\subsubsection{Proof}

\paragraph{Forward Direction: \( user \land object1 = object1 \implies user^C \bullet object1 = 0 \)}

Assume \( user \land object1 = object1 \).

\begin{itemize}
    \item This means every element in object1 is also in user. Formally, \( \forall x (x \in object1 \implies x \in user) \).
    \item The complement \( user^C \) consists of elements not in user. Therefore, \( x \notin user^C \) for any \( x \in object1 \).
    \item Since no element of object1 is in \( user^C \), the intersection of \( user^C \) and object1 is empty: \( user^C \cdot object1 = 0 \).
\end{itemize}

\paragraph{Backward Direction: \( user^C \bullet  object1 = 0 \implies user \land object1 = object1 \)}

Assume \( user^C \bullet object1 = 0 \).

\begin{itemize}
    \item This means there is no element in object1 that is also in \( user^C \). Formally, \( \forall x (x \in object1 \implies x \notin user^C) \).
    \item Since \( x \notin user^C \) for any \( x \in object1 \), it follows that \( x \in user \).
    \item Therefore, every element in object1 is in user, which means \( user \land object1 = object1 \).
\end{itemize}

\subsubsection{Definitions and Initial Setup for $ user^C \bullet object2 = 1 $}

\begin{itemize}
    \item \( user^C \): The complement of user, representing all elements not in user.
    \item Dot Product: In this context, it is a boolean operation indicating the intersection. For boolean values, \( A \bullet B = 1 \) if both \( A \) and \( B \) are 1; otherwise, it is 0.
    \item Conjunction: \( \land \) represents the logical AND operation.
\end{itemize}

\subsubsection{Proof Steps for - $ user^C \bullet object2 = 1 $}

We need to show that:
\[ user \land object2 \neq object2 \iff user^C \bullet object2 = 1 \]

\subsubsection{Proof}

\paragraph{Forward Direction: \( user^C \bullet object2 = 1 \implies user \land object2 \neq object2 \)}

Assume \( user^C \bullet object2 = 1 \).

\begin{itemize}
    \item This means there exists at least one element in object2 that is also in \( user^C \). Formally, \( \exists x (x \in object2 \land x \in user^C) \).
    \item Since \( x \in user^C \), it follows that \( x \notin user \).
    \item Therefore, not every element in object2 is in user, which means \( user \land object2 \neq object2 \).
\end{itemize}

\paragraph{Backward Direction: \( user \land object2 \neq object2 \implies user^C \bullet object2 = 1 \)}

Assume \( user \land object2 \neq object2 \).

\begin{itemize}
    \item This means there exists at least one element in object2 that is not in user. Formally, \( \exists x (x \in object2 \land x \notin user) \).
    \item Since \( x \notin user \), it follows that \( x \in user^C \).
    \item Therefore, there is at least one element in object2 that is in \( user^C \), which means \( user^C \cdot object2 = 1 \).
\end{itemize}

\subsubsection{Conclusion}

By proving both directions for both expressions, we have shown the equivalence:
\[ user \land object1 = object1 \iff user^C \bullet object1 = 0 \]
and
\[ user \land object2 \neq object2 \iff user^C \bullet object2 = 1 \]
holds. A valid test for Dominance can be performed by a dot product between the User and the Object. 

\section{IBAC Security Label Encoding for Orthogonality and Dominance implemented
as Aggregate functions for massive performance enhancement}

In the last section, we have shown that the dot product can be used as a test of dominance. The dot product is an interesting construct; there exists an algebraic
and a geometric definition. \par

\medskip

We discussed earlier the dot product and the use of it for dominance from an algebraic perspective. The question then to ask; is it possible to look at its geometric definition in the context of a trusted system, i.e., whether we can also use it for testing for dominance?

 \medskip 

${a\bullet b}=\bigparallel a\bigparallel\:\bigparallel b\bigparallel\:\cos(\theta)$ 

 \medskip 

On the \textbf{left-hand} side of the equal symbol, there is an operator
that algebraically operates on each of the elements of the two vectors
to test for orthogonality. We looked at this earlier from a boolean perspective. The 'left-hand' side, in a sense, is in
the world of a vector space for the security 'domain of discourse.' 

 \medskip 

On the \textbf{right-hand} side of the equal symbol, there is its
geometric interpretation as an angle between two absolute numerical
values, each representing the vector lengths of both a and b. 'The
'right-hand side' of the equation is the length, the \textbf{aggregate}
value that creates the geometric interpretation of orthogonality,
or of being perpendicular. 

 \medskip 

Let's use the right-hand side geometric definition insight of the
dot product as an analogy, to see if other mathematical \textbf{aggregate}
\textbf{structures} exist that enable a way of testing for dominance.
I.e., by a \textbf{right-hand side } test for orthogonality, between
the two vectors $a\,_{User}$ and $b_{\,Object}$. 

 \medskip 

The research carried out over the years has led to the development
of two orthogonal encoding methods that are analogous to the geometric
definition of orthogonality; both have a \textbf{left and right-hand
side} to the definition. 

 \medskip 

Before explaining the encoding methods, it is worth knowing that Donald
Knuth \cite{key-6} uses the perpendicular symbol between two integers
to signify that they are relatively prime, $\mathbf{a}\perp\mathbf{b}$.
Knuth sees two integers that are relative prime to each other as also
orthogonal under the operation of multiplication.  

 \medskip 

Taking the following three prime numbers in a set as an example $\{7,11,13\}$.
Any combination of the three primes, only used once, is a unique number
under the $\prod$ product operation, i.e. all combinations of the
power set are orthogonal to each other under multiplication. 

 \medskip 

Integers can also be orthogonal to each other under the operation
of $\sum$ addition. The unix / linux chmod operation uses this exact
concept with the various combinations taken from the set $\{2^{0},2^{1},2^{2}\}$
, for the read, write, execute combinations for file access control.

 \medskip 

 \medskip 

Extending this further, taking randomly any combinations of unique
integers from a set that are exponents to some common base, the combination
under addition will be unique and they will behave as $\mathbf{\mathbf{\mathbf{a}}}\perp\mathbf{b}$ 

 \medskip 

Using the following set of three exponents as an example $\{3^{1},3^{5},3^{7}\}$
any combination of the three, only used once, is a unique number under
the sum $\sum$ operation. I.e., just like the unix chmod operation,
the various combinations are orthogonal to each other. 

 \medskip 

What is interesting is that when using the developed \textbf{addition}
and \textbf{multiplication} orthogonal encoding schemes, like the
dot product, there is a left and right hand side, respectively a vector
and and aggregate of the vector:
\begin{itemize}
\item Under addition, the \textbf{left hand} side of the vector $[3^{1},3^{5},3^{7}]$
becomes the \textbf{right hand} aggregate of $3^{1}+3^{5}+3^{7}=2433$ 
\item Under multiplication, the \textbf{left hand} side vector $[7,11,13]$
becomes the \textbf{right hand} aggregate of $7*11*13=1001$ 
\end{itemize}
The benefit of these orthogonal encoding schemes is the problems of
bit-vectors regarding trusted systems as described in prior section
4 cease to exist.

\section{IBAC Orthogonal Addition encoding dealing with the problem of sparse arrays}

\subsection{The mechanics of addition encoding}

Let us look in detail at the first Orthogonal Encoding method, developed
as an alternative to bit-vectors, integer exponents to some common
base.  

 \medskip 

Continuing the discussion using the security example from above, the
set of labels for the universe of security discourse is defined as
follows:  

 \medskip 

\begin{tabular}{|l||l|}
\hline 
\textbf{Security Label semantic} & $integerToken$\tabularnewline
\hline 
Top secret & $3^{0}$\tabularnewline
\hline 
Secret & $3^{1}$\tabularnewline
\hline 
Protected  & $3^{2}$\tabularnewline
\hline 
Public & $3^{3}$\tabularnewline
\hline 
GCHQ  & $3^{4}$\tabularnewline
\hline 
MI5 & $3^{5}$\tabularnewline
\hline 
MI6 & $3^{6}$\tabularnewline
\hline 
\end{tabular} \par
\textbf{figure}-5: A set of exponents for the universe of discourse 

 \medskip 

Orthogonal addition encoding is defined as $N=\sum_{k=0}^{k=n}(C_{k}b^{k})$,
such that $N\in\mathbb{Z^{\text{+*}}}$, $C_{k}\in\{0,1\}$ and $b\geqq2$

 \medskip 

 \medskip 

\textbf{P\_rel:} 
Orthogonality still applies as a test of dominance on the left-hand-side
of the dot product, so using the orthogonal complement $User^{c}$
for each $User$ results in:  

 \medskip 

\begin{tabular}{|l||l||l||l|l|}
\hline 
\textbf{Access Granted} & $user$ & $user^{C}$ & $object1$ & $user^{C}\bullet object1$\tabularnewline
\hline 
Top secret & 0 & $3^{0}$ & 0 & 0\tabularnewline
\hline 
Secret & $3^{1}$ & 0 & $3^{1}$ & 0\tabularnewline
\hline 
Protected  & $3^{2}$ & 0 & 0 & 0\tabularnewline
\hline 
Public & $3^{3}$ & 0 & 0 & 0\tabularnewline
\hline 
GCHQ  & 0 & $3^{4}$ & 0 & 0\tabularnewline
\hline 
MI5 & $3^{5}$ & 0 & $3^{5}$ & 0\tabularnewline
\hline 
MI6 & $3^{6}$ & 0 & 0 & 0\tabularnewline
\hline 
\multicolumn{1}{|l|}{$\mathbf{\sum exponents\Rightarrow=\perp}$} & \multicolumn{1}{l}{} & \multicolumn{1}{l}{} &  & \textbf{0}\tabularnewline
\hline 
\end{tabular} \par
\textbf{figure}-6: exponent dominance operation of: 
 $((Secret,\{MI5,MI6\}),(Secret,\{MI5\}))\in dominates$ 

 \medskip 

\textbf{N\_rel:} 
and also where dominance does not apply such that access is denied
results in : 

 \medskip 

\begin{tabular}{|l||l||l||l|l|}
\hline 
\textbf{Access Denied} & $user$ & $user^{C}$ & $object1$ & $user^{C}\bullet object1$\tabularnewline
\hline 
Top secret & 0 & $3^{0}$ & 0 & 0\tabularnewline
\hline 
Secret & $3^{1}$ & 0 & $3^{1}$ & 0\tabularnewline
\hline 
Protected  & $3^{2}$ & 0 & 0 & 0\tabularnewline
\hline 
Public & $3^{3}$ & 0 & 0 & 0\tabularnewline
\hline 
GCHQ  & 0 & $3^{4}$ & $3^{4}$ & $3^{8}$\tabularnewline
\hline 
MI5 & $3^{5}$ & 0 & 0 & 0\tabularnewline
\hline 
MI6 & $3^{6}$ & 0 & $3^{6}$ & 0\tabularnewline
\hline 
\multicolumn{1}{|l|}{$\mathbf{\sum exponents\Rightarrow\neq\perp}$} & \multicolumn{1}{l}{} & \multicolumn{1}{l}{} &  & \textbf{$\mathbf{3}^{\mathbf{8}}$}\tabularnewline
\hline 
\end{tabular} \par
\textbf{figure}-7: exponent dominance operation of: 
 $((Secret,\{M15,MI6\}),(Secret,\{GCHQ,MI6\}))\notin dominates$

\subsection{Aggregate Value of sum of exponents - the right hand side}

The structure of the addition aggregate is of the form:

$A\in\mathbb{Z^{\text{+}\ast}},\,\:A=b^{n}+(b^{n-1}\lor0)+(b^{n-2}\lor0)...+(b^{0}\lor0)$
where $b$ is some radix value and $n$ is the largest integer index
in the set of semantics being encoded. 

 \medskip 

The addition aggregate function is analogous to $\bigparallel a\bigparallel\:\bigparallel b\bigparallel\:\cos(\theta)$.
The \textbf{left hand side} of the equal sign is the vector, i.e.
a vector list of exponents of some common base. The \textbf{right
hand side} of the equal sign is the encoding summation of all the
exponents of some common base. 
$[b^{n},(b^{n-1}\lor0),(b^{n-2}\lor0),...(b^{0}\lor0)]_{left\:hand\:side}\;where\;b^{n}+(b^{n-1}\lor0)+(b^{n-2}\lor0)...+(b^{0}\lor0)_{right\:hand\:side}\;is\;the\;aggregate$

 \medskip 

 \medskip 

The right-hand side aggregate function calculates the 'addition' orthogonal
value and provides:
\begin{itemize}
\item The aggregate value, such that when the value is decoded using the
deconstruct function below (Algorithm 1), a list of exponents that
were initially encoded is returned.
\item When testing for dominance, a variation of Algorithm 1, (i.e., Algorithm
2) Compares orthogonality for the various vector elements as they
are decoded. 
\end{itemize}
In the case of the figure-6, $((Secret,\{MI5,MI6\}),(Secret,\{MI5\}))\in dominates$
: 
$([3^{1}+3^{2}+3^{3}+3^{5}+3^{6}]_{user}=1011{}_{user},\;\;[3^{1}+3^{5}]_{object}=246_{object})\in dominates$

 \medskip 

 \medskip 

In the case of the figure-7, $((Secret,\{M15,MI6\}),(Secret,\{GCHQ,MI6\}))\notin dominates$
: 
($[3^{1}+3^{2}+3^{3}+3^{5}+3^{6}]_{user}=1011{}_{user},\;\;[3^{1}+3^{4}+3^{6}]_{object}=813_{object})\notin dominates$ 

\subsection{Dealing with sparse arrays of bit-vectors }

We have discussed the importance of orthogonality on the left-hand
side as an alternative test for dominance, but why the right-hand
side aggregate representation?  

 \medskip 

It turns out that most trusted systems are practical problems involving
sparse arrays. This is a subtle but important point. Many organisations
have many security markings available, but only a few are ever combined
at any point in time for combinations of $(User,Object)$. If implemented
as bit-vectors then there are many 0's to manage.  

 \medskip 

If an aggregate function that keeps the notion of orthogonality can
be found, such as 'addition' of exponents to some base, then it is
not required to deal with the zero positions of the vectors. The sparse
array problem goes away. 

 \medskip 

All operations, comparative to bit-vectors on average are simplified
and are $\frac{N}{2}$, as the zero positions don't form part of the
encoding. I.e. all operations on an aggregate encoding holding on
average $\frac{N}{2}$ elements and will then provide a performance
gain with a reduction in complexity.  

 \medskip 

Avoiding the problem of sparse array inefficiencies then makes the
maintenance of the aggregate security tokens very simple to manage.
In relation to the Orthogonal Addition encoding, simple operations
of addition and subtraction are used to maintain security tokens.
Adding and subtracting the various semantics as required, which provides
the abstraction of the actual bit-vectors. 

\subsection{Aggregate dominance }

It is one thing to encode the security semantics; it is another to
decode the semantics to test for dominance efficiently. 

 \medskip 

As it turns out, there is a simple iterative function for decoding
in any base that can be slightly modified to then test for dominance. 

 \medskip 

The iterative function for decoding orthogonal addition is based on
the following numerical observation: 

 \medskip 

$x_{n+1}=x_{n}-base^{floor(log_{base}(x_{n}))}$ such that $x_{n+1}\geq1$ 

 \medskip 

The simplified complexity of the above iteration is related to $\frac{N}{2}$
on average as the semantics that were zero were never encoded, and
are therefore never encountered when decoded.  

 \medskip 

For example the object $((Secret,\{MI5\})_{object}$ aggregate of
$3^{1}+3^{5}=246_{10}$ will decode as follows: 

 \medskip 

$x_{n+1}=x_{n}-base^{floor(log_{base}(x_{n}))}$ 
$3_{10}=246_{10}-3^{floor(log_{3}(246_{10})}=246_{10}-3^{5}$, i.e.,
the first index is 5 
$0=3_{10}-3^{floor(log_{3}(3_{10})}=3_{10}-3^{1}$, i.e., the second
index is 1 
An array of indices $[5,1]_{3}$ is returned  

 \medskip 

The full code for decoding or deconstruct, i.e. the name of the algorithm,
is as follows: 

 \medskip 

\begin{algorithm}[h]
\caption{deconstruct\protect\textsubscript{}(\textbf{input}: aInteger, base)}

$iArray\leftarrow[]$

$temparyInteger\leftarrow aInteger$

\textbf{while} $temparyInteger\geq1$ \textbf{do}

\qquad{}$index\leftarrow\mathbf{floor(log_{base}}(temparyInteger))$

\qquad{}$iArray.append(index)$

\qquad{}$temporaryInteger\leftarrow temporaryInteger-base^{index}$

\textbf{return} $iArray$
\end{algorithm}

 \medskip 

As long as the indices used for encoding are unique, never repeated,
this algorithm's complexity is $\mathcal{O}2log\frac{N}{2}$ removing the constant $\mathcal{O}log N $, on
average only $\frac{N}{2}$ operations are required. The dominance
relation itself is almost as simple: 

 \medskip 

\begin{algorithm}[h]
\caption{dominance\protect\textsubscript{}(\textbf{input}: aInteger, bInteger,
base)}

$temparyIntegerA\leftarrow aInteger$

$temparyIntegerB\leftarrow bInteger$

\textbf{while} $temparyIntegerA\geq1$ \textbf{do}

\qquad{}$indexA\leftarrow\mathbf{floor(log_{base}}(temparyIntegerA))$

\qquad{}\textbf{while} $temparyIntegerB\geq1$ \textbf{do}

\qquad{}\qquad{}$indexB\leftarrow\mathbf{floor(log_{base}}(temparyIntegerB))$

\qquad{}\qquad{}\textbf{if} $indexA>indexB$

\qquad{}\qquad{}\qquad{}\textbf{break from loop}

\qquad{}\qquad{}\textbf{else if} $indexA=indexB$

\qquad{}\qquad{}\qquad{}$temporaryIntegerB\leftarrow temporaryIntegerB-base^{indexB}$

\qquad{}\qquad{}\qquad{}\textbf{break from loop}

\qquad{}\qquad{}\textbf{else} return $-1$ \#dominate relation is
false between aInteger and bInteger

\qquad{}$temporaryIntegerA\leftarrow temporaryIntegerA-base^{indexA}$

\textbf{return} $0$ \#dominance relation holds between aInteger and
bInteger
\end{algorithm}

 \medskip 

The expectation of the inner loop when testing for dominance (Algorithm
2) between $aInteger,bInteger$, is that the control flow will break
on each iteration of the inner loop for dominance to hold  

 \medskip 

As long as each index is used only once (the critical assumption of
the working dominance algorithm) , regardless of the base, this algorithm
is extremely fast, it's complexity the same as Algorithm 1 $\mathcal{O}log N $

 \medskip 

 \medskip 

Using the knowledge that the index sum operation $\sum$ is a combination
of orthogonal integers with the ability to quickly decode the aggregate
results in a useful security abstraction of the bit-vector.  

 \medskip 

An abstraction that can be applied to any layer in the technology
stack to provide authorisation for access: 
\begin{itemize}
\item Rows in the RDBMS
\item Records in a file System
\item Documents in noSQL database
\item messages in a message broker
\item Even the OS file system
\end{itemize}
As interesting as Orthogonal addition encoding is, the gold standard
in terms of performance from the research was when testing for dominance
using the orthogonal prime number encoding scheme.

\section{IBAC Orthogonal Prime Number encoding}

\subsection{The mechanics of Products of primes encoding}

The second method developed for trusted systems was to utilize prime
numbers. Continuing again with both Donald Knuth's use of the perpendicular
symbol for relative primes, $\mathbf{a}\perp\mathbf{b}$ and the previous
security example from above; the set of labels for the universe of
security discourse can be defined as follows: 

 \medskip 

\begin{tabular}{|l||l|}
\hline 
\textbf{Security Label semantic} & $integerToken$\tabularnewline
\hline 
Top secret & $3$\tabularnewline
\hline 
Secret & $5$\tabularnewline
\hline 
Protected  & $7$\tabularnewline
\hline 
Public & $11$\tabularnewline
\hline 
GCHQ  & $13$\tabularnewline
\hline 
MI5 & $17$\tabularnewline
\hline 
MI6 & $19$\tabularnewline
\hline 
\end{tabular}  \par
\textbf{figure} 8: A set of prime numbers for the universe of discourse
such that each prime is also orthogonal (perpendicular) to each other. 

 \medskip 

The prime number encoding method is the gold standard method, testing
for the dominance relation is trivial once the orthogonal encoding
is in-place 

 \medskip 

Orthogonal multiplication encoding is defined as $N=\prod_{k=1}^{k=n}(P_{k})$,
such that $N\in\mathbb{Z^{\text{+*}}}$, $P_{k}\in\{primes\}$ and
$P_{k}\geqq2$  

 \medskip 

\textbf{P\_rel:} 

 \medskip 

Orthogonality still applies as a test of dominance using the dot product
between the $User^{C}\:and\;Object$ :  

 \medskip 

 \medskip 

\begin{tabular}{|l||l||l||l|l|}
\hline 
\textbf{Access Granted} & $user$ & $user^{C}$ & $object1$ & $user^{C}\bullet object1$\tabularnewline
\hline 
Top secret & 0 & $3$ & 0 & 0\tabularnewline
\hline 
Secret & $5$ & 0 & $5$ & 0\tabularnewline
\hline 
Protected  & $7$ & 0 & 0 & 0\tabularnewline
\hline 
Public & $11$ & 0 & 0 & 0\tabularnewline
\hline 
GCHQ  & 0 & $13$ & 0 & 0\tabularnewline
\hline 
MI5 & $17$ & 0 & 17 & 0\tabularnewline
\hline 
MI6 & $19$ & 0 & 0 & 0\tabularnewline
\hline 
\multicolumn{1}{|l|}{$\mathbf{\sum primes\Rightarrow=\perp}$} & \multicolumn{1}{l}{} & \multicolumn{1}{l}{} &  & \textbf{0}\tabularnewline
\hline 
\end{tabular}  \par
\textbf{figure} 9: prime number dominance operation of: 
 $((Secret,\{MI5,MI6\}),(Secret,\{MI5\}))\in dominates$ 

 \medskip 

\textbf{N\_rel:} 

 \medskip 

and also where dominance does not apply using the dot product such
that access is denied: 

 \medskip 

\begin{tabular}{|l||l||l||l|l|}
\hline 
\textbf{Access Denied} & $user$ & $user^{C}$ & $object1$ & $user^{C}\bullet object2$\tabularnewline
\hline 
Top secret & 0 & $3$ & 0 & 0\tabularnewline
\hline 
Secret & $5$ & 0 & $5$ & 0\tabularnewline
\hline 
Protected  & $7$ & 0 & 0 & 0\tabularnewline
\hline 
Public & $11$ & 0 & 0 & 0\tabularnewline
\hline 
GCHQ  & 0 & $13$ & $13$ & 169\tabularnewline
\hline 
MI5 & $17$ & 0 & 0 & 0\tabularnewline
\hline 
MI6 & $19$ & 0 & $19$ & 0\tabularnewline
\hline 
\multicolumn{1}{|l|}{$\mathbf{\sum primes\Rightarrow\neq\perp}$} & \multicolumn{1}{l}{} & \multicolumn{1}{l}{} &  & 169\tabularnewline
\hline 
\end{tabular}  \par
\textbf{figure} 10: prime number dominance operation of: 
 $((Secret,\{M15,MI6\}),(Secret,\{GCHQ,MI6\}))\notin dominates$

\subsection{Aggregate Value of product of primes - the right hand side}

The rationale for the orthogonal product of prime encoding as the
gold standard in the aggregate is the test for dominance reduces down
to a simple modulo test between the two carefully encoded security
tokens $(User,Object)$. 

 \medskip 

For encoding the dominates relation figure 9 above $((Secret,\{MI5,MI6\}),(Secret,\{MI5\}))\in dominates$ 

 \medskip 

Encoding the $User$ Orthogonal token$(Secret,\{MI5,MI6\})$ is 
$5*7*11*17*19=124355$ 

 \medskip 

Encoding of the $Object$ Orthogonal token $(Secret,\{MI5\}$ is also
straight forward 
 $5*17=85$ 

 \medskip 

A test for dominance, $(User,Object)\in dominate$ is as simple as
testing the modulo of $(User,Object)==0$: 
 $0==124355\:mod\:85$ 

 \medskip 

For encoding the \textbf{not} dominates relation figure 10 above,
$((Secret,\{M15,MI6\}),(Secret,\{GCHQ,MI6\}))\notin dominates$ : 

 \medskip 

Encoding the $User$ Orthogonal token$(Secret,\{MI5,MI6\})$ is again 
$5*7*11*17*19=124355$ 

 \medskip 

Encoding of the $Object$ Orthogonal token $(Secret,\{GCHQ,MI6\}$
is also straight forward 
 $5*13=65$ 

 \medskip 

A test for dominance that does not hold for the relation, $(User,Object)\notin dominate$
is where the remainder of the modulo function is greater than zero: 
 $45==124355\:mod\:65$ 

\subsection{Orthogonal Encoding is an abstraction from actual bits and requires
more memory}

In any new encoding scheme, there are pro's and con's; you don't achieve
the advantage of a new system that is difficult to hack and is highly
maintainable without some disadvantage. One cannot escape the digital
universe, the binary machine that executes code. The orthogonal multiplication
and addition 'aggregate codes' reduce down to bit-vectors. The execution is on abstractions of bit vector encoding, but storage of abstractions are bit-vectors. \par

 \medskip 

As the example below demonstrates, the abstractions of the multiplication
and addition orthogonal encoding scheme (addition other than base
2) reduces to a bit-vector size that requires more space than encoding
in strictly a binary method. 

 \medskip 

Assuming the three semantics A, B and C and the various encoding from Section 5:
\begin{itemize}
\item Using a strictly \textbf{binary} encoding scheme, the three codes: $[A=001,B=010,C=100]$
when combined under binary addition will be $[111]$, i.e. a member
of the set of bits (1s and 0s) up to the Mersenne number $2^{3}-1$
\item Using the \textbf{orthogonal addition} encoding scheme of radix 3, three codes:
$[A=3^{0},B=3^{1},C=3^{2}]$ when combined under Orthogonal addition
$[3^{0}+3^{1}+3^{2}]$ will be $[1101]$, i.e. a member of a set of
bits (1s and 0s) up to the Mersenne number $2^{4}-1$
\item Using the \textbf{orthogonal multiplication} scheme of primes, three codes:
$[A=3,B=5,C=7]$ when combined under Orthogonal multiplication $[3*5*7]$
will be $[1101001]$, i.e. a member of a set of bits (1s and 0s) up
to the Mersenne number $2^{7}-1$
\end{itemize}

The point being made is that the benefits of the multiplication and addition
orthogonal encoding are offset by the requirements of additional memory
that is then required, comparative to a strictly binary encoding scheme,
to store the orthogonal encoded integers. \par
\medskip

Having made the increased storage point, for the sake of completeness,  for practical purposes; (1) with modern cloud platforms and modern data centres, storage cost keeps reducing, and (2) we are talking about the storage of Security Token tags (the meta data) which is minuscule in comparison to the actual data storage cost.

\section{The application of IBAC for AI and AI Automation }

\section{Discussion of benefits for Orthogonal Encoding using IBAC}

For both orthogonal encoding methods we have described (1) \textbf{ addition of indexes} 
of some radix and (2) \textbf{products of primes} , the utility gained
by the dominance relation is related to orthogonality. Not so
much the Algebraic definition of the dot product but the equal of an aggregated Geometric
definition based on some aggregated function which deals with both the
problems of sparse arrays as well as providing superior performance when filtering data. 

\subsection{Benefits of Orthogonal Security Token encoding}

The practical benefits of both the orthogonal encoding methods is provided by the leverage of the aggregate functions and despite a small increase in memory for the orthogonal encoding abstraction, the benefits in contrast to bit-vectors include the following:
\begin{itemize}
\item Simplicity
\item Flexibility
\item Highly performant 
\item Ubiquitous
\item Interoperable 
\end{itemize}

\subsection{Simplicity}

The algorithms and technology developed are both: 
\begin{enumerate}
\item Simple to implement and 
\item Simple to deploy. 
\end{enumerate}
From the governance perspective simplicity means: 
\begin{itemize}
\item It is straightforward to understand how the security policy has been
implemented and 
\item The assurance of a correctly working policy implemented is also straightforward
to verify. 
\end{itemize}
The simplicity of implementation leads to the ease of any business
governance process to continually audit the verification of a correctly
working piece of software and its ongoing integrity and compliance
over time. 

 \medskip 

In addition, the simplicity of implementation and the encoding allows
ease of communication and discourse of the actual policy implemented
to the executive branch, who may not be technical. 

 \medskip 

The \textquotedblleft Declarative Nature\textquotedblright{} of the
approach also enhances simplicity. \textquoteleft \textbf{Declarative}\textquoteright{}
meaning the ability to declare what the policy does, without requiring
to hard code any security rules and being able to rely on a verifiable
but abstract simple high-level modulo (in the case of prime number
encoding ) function to implement the dominance relation. This only
adds to the simplicity of audit and compliance of the control. 

\subsection{Flexibility}

The declarative simplicity of the algorithms means the algorithms
can be flexibly applied. 

 \medskip 

For example, if using a rules-based approach implemented directly
using bit-vectors, the state of any 'security rule' for a given security
token would be stored as bits in the vector, where each bit 0 or 1
holds the meaning of applying or not applying a particular security
label.  

 \medskip 

The bit-vector requires each bit of 0 or 1 of the token to be individually
managed, as each bit designates if a particular security label (semantic)
is applied or is not applied to the token. There is also a limit to
the bit-vector size itself. All are adding complexity to when the
need arises to increase the size and maintain the bit-vector. 

 \medskip 

The abstraction of the bit-vector, using 'prime numbers' or 'exponents
to some base', requires an integer product or integer sum and to remove
or add a security label requires only the simple arithmetic operators
of division in the case of products of primes or subtraction in the
case of sums of exponents to some base. 

 \medskip 

To remove $3^{5}$ from the user label $[3^{1}+3^{2}+3^{3}+3^{5}+3^{6}]_{user}=1011$,
all that is required is to subtract $1011-3^{5}=768=[3^{1}+3^{2}+3^{3}+3^{6}]_{user}$-
. Note that there is no need to manipulate a bit-vector to set 1 or
0. 
To remove 17 from the user label $5*7*11*17*19=124355_{user}$, all
that is required is to divide $\frac{124355}{17}=7315_{user}=5*7*11*19$ 

 \medskip 

There are two crucial implications of this approach that together
create great flexibility for managing security tokens: 
\begin{itemize}
\item The Security labels as prime numbers or indexes to some base that
are not applied in a token do not exist as part of the product or
sum for a specific encoded token, and it then follows, components
of a product or sum that are absent, are not required to be managed. 
\item There is also no notion of bit-vector size or upper limit of bits,
of the number representing security labels (semantics). There exists
an infinite theoretical capacity in terms of multiplication or summation,
limited only by the underlying machine technology to process big integers.
\end{itemize}
The flexibility aspects of the technology go hand in hand with being
able to make the technology ubiquitous and interoperable. 

\subsection{Highly performant}

The declarative simplicity of the prime number or exponent algorithms,
means the algorithms are fast, highly performant. The products of
primes algorithm have been implemented on a 400 million row test Teradata
Databases with negligible performance loss using complex joins between
tables. 

 \medskip 

The alternative rule-based coded approach on Teradata compared at
the same time of testing suffered an exponential performance loss
with joins, whereas the bit vector abstraction, the product of primes
algorithm, incurred a small linear performance degradation with join
complexity. 

 \medskip 

The reason for the better orthogonal encoding performance on the RDBMS
is the performance hit is not taken on utilizing, decoded the security
policy (i.e., the READ), but on the initial encoding process, the
INSERT or UPDATE. I.e. the performance benefit gain from utilizing
the novel algorithms is by moving the computational complexity away
from the security policy (READ), which typically is carried out many
thousands of times once the data is operational, moved to the orthogonal
encoding mechanism which is done once at set up (INSERT) and on maintenance
(UPDATE). 

 \medskip 

The additional reason for the performance gain for the prime number
encoding method is that a simple \textbf{modulo function} is used
for the implementation of dominance. The benchmark testing on 1000,
10000, 100000 and 1000000 reads comparing bit-vectors $(A\nsubseteq B)\:\:iff\:\:(A\cap B\neq A)$
and modulo$(A\nsubseteq B)\:\:iff\:\:(A\:mod\:B\neq0)$ indicates
the bit-vector dominance operation AND comparative to an orthogonal
prime number algorithm using modulo is typically very close in performance
over thousands of reads with possibly a more comprehensive performance
gain by using modulo when the arrays are sparse.  

 \medskip 

The User bit-vector Token set as $[1011111111]$ and Object bit-vector
Token as $[1110111111]$ 
The User Orthogonal Prime Token set as $2*5*7*11*13*17*19*23*31*61$
and Object Orthogonal Prime Token as 2{*}3{*}5{*}13{*}17{*}19{*}23{*}31{*}37{*}61 
Note that for both the bit-vector and Orthogonal Prime case $(User,Object)\notin dominate$ 

 \medskip 

Post testing, the bit-vector AND operation, and the product of primes
modulo operation were on par. At one level it was surprising the modulo
operation performed as well as the bit AND operation, but at another
level, if there is one atomic operation that has been well optimised
over the years in both hardware and the OS for the express purpose
of encryption/decryption, it is the OS modulo operation.  

 \medskip 

The tables below are a summary of the dominance relation testing.
The first two tables are the results of testing modulo vs. bit-vector,
where the dominance relation does not hold. The executed python code
for the testing ran on an Intel Core i9, 32 gig of ram, running within
the Komodo IDE. The code iterates over the one predicate, both bit-vector,
and products of primes, timing the corresponding number of reads,
utilising dominance. 

 \medskip 

\begin{tabular}{|l|l|l|}
\hline 
\textbf{\scriptsize{}number }{\scriptsize{}of reads} & \textbf{\scriptsize{}time Seconds}{\scriptsize{} }\textsubscript{{\scriptsize{}10 bits AND}} & \textbf{\scriptsize{}time Seconds}{\scriptsize{} }\textsubscript{{\scriptsize{}10 Primes modulo}}\tabularnewline
\hline 
\hline 
{\scriptsize{}1,000} & {\scriptsize{}0.000158071517944} & {\scriptsize{}0.000159978866577}\tabularnewline
\hline 
{\scriptsize{}10,000} & {\scriptsize{}0.0013210773468} & {\scriptsize{}0.00133299827576}\tabularnewline
\hline 
{\scriptsize{}100,000} & {\scriptsize{}0.0127429962158} & {\scriptsize{}0.0128610134125}\tabularnewline
\hline 
{\scriptsize{}1,000,000} & {\scriptsize{}0.109389066696} & {\scriptsize{}0.097678899765}\tabularnewline
\hline 
\end{tabular} 

 \medskip 

\begin{tabular}{|l|l|}
\hline 
\textbf{\scriptsize{}number }{\scriptsize{}of reads} & \textbf{\scriptsize{}time Seconds}{\scriptsize{}}\textsubscript{{\scriptsize{}primes modulo}}{\scriptsize{}/
}\textbf{\scriptsize{}time Seconds}{\scriptsize{}}\textsubscript{{\scriptsize{}bit AND}}\tabularnewline
\hline 
\hline 
{\scriptsize{}1,000} & {\scriptsize{}modulo time, 101.206636501 \% of bit-vector AND operation}\tabularnewline
\hline 
{\scriptsize{}10,000} & {\scriptsize{}modulo time, 100.902364194 \% of bit-vector AND operation}\tabularnewline
\hline 
{\scriptsize{}100,000} & {\scriptsize{}modulo time, 100.926133812 \% of bit-vector AND operation}\tabularnewline
\hline 
{\scriptsize{}1,000,000} & {\scriptsize{}modulo time, 89,2949384387 \% of bit-vector AND operation}\tabularnewline
\hline 
\end{tabular} 

 \medskip 

 \medskip 

The second pair of tables is the result of testing of dominance on
a sparse array. The $Object$ bit-vector token is initialised to $[1000000001]$,
with 8 zero bits in the bit-vector. Also the $Object$ orthogonal
prime number token is initialised to $2*61$. The tables below are
a summary of the dominance relation testing for the enoding that is
a sparse array of bits; testing, modulo vs. bit-vector, where the
dominance relation does hold. Again, the executed python code was
running on an Intel Core i9, 32 gig of ram, running within the Komodo
IDE. The code iterates over the one predicate, both bit-vector, and
products of primes, timing the corresponding number of reads, utilising
dominance. 

 \medskip 

\begin{tabular}{|l|l|l|}
\hline 
\textbf{\scriptsize{}number }{\scriptsize{}of reads} & \textbf{\scriptsize{}time Seconds}{\scriptsize{} }\textsubscript{{\scriptsize{}10 bits AND}} & \textbf{\scriptsize{}time}{\scriptsize{} }\textbf{\scriptsize{}Seconds}{\scriptsize{}}\textsubscript{{\scriptsize{}10 Primes modulo}}\tabularnewline
\hline 
\hline 
{\scriptsize{}1,000} & {\scriptsize{}0.000158071517944} & {\scriptsize{}0.000164031982422}\tabularnewline
\hline 
{\scriptsize{}10,000} & {\scriptsize{}0.00168395042419} & {\scriptsize{}0.00150394439697}\tabularnewline
\hline 
{\scriptsize{}100,000} & {\scriptsize{}0.014899969101} & {\scriptsize{}0.0139911174774}\tabularnewline
\hline 
{\scriptsize{}1,000,000} & {\scriptsize{}0.116893053055} & {\scriptsize{}0.114114999771}\tabularnewline
\hline 
\end{tabular} 

 \medskip 

\begin{tabular}{|l|l|}
\hline 
\textbf{\scriptsize{}number }{\scriptsize{}of reads} & \textbf{\scriptsize{}time Seconds}{\scriptsize{}}\textsubscript{{\scriptsize{}primes modulo}}{\scriptsize{}/
}\textbf{\scriptsize{}time Seconds}{\scriptsize{} }\textsubscript{{\scriptsize{}bits AND}}\tabularnewline
\hline 
\hline 
{\scriptsize{}1,000} & {\scriptsize{}modulo time, 103.770739065 \% of bit-vector AND operation}\tabularnewline
\hline 
{\scriptsize{}10,000} & {\scriptsize{}modulo time, 89.3104912927 \% of bit-vector AND operation}\tabularnewline
\hline 
{\scriptsize{}100,000} & {\scriptsize{}modulo time, 93,.900312025\% of bit-vector AND operation}\tabularnewline
\hline 
{\scriptsize{}1,000,000} & {\scriptsize{}modulo time, 97.623423111\% of bit-vector AND operation}\tabularnewline
\hline 
\end{tabular} 

 \medskip 

 \medskip 

It would seem that when dominance is applied to sparse arrays, there
may be a slight improvement when using the modulo function rather
than the bit AND operation. The results are far from conclusive, and
it is suggested that full benchmarking is probably a good area of
further research.  

 \medskip 

However, what can be said for certain is the 'modulo' and bit-vector
'AND' operations are on par in terms of performance. 

\subsection{Ubiquitous}

The algorithms and technology developed are so flexible and straightforward,
it can be ubiquitously applied across the entire fabric of an organisation\textquoteright s
technology stack.  

 \medskip 

Being ubiquitous means the 'one' organisation\textquoteright s security
policy can be consistently implemented across all components of all
systems for that organisation and between actual systems of various
organisations. 

 \medskip 

The orthogonal token-based trusted systems implementation of a security
policy, either product of 'prime numbers' or sums of 'exponents to
some base', can be ubiquitously applied to the following layers in
the technology fabric: 
\begin{enumerate}
\item Application 
\item Database 
\item File system 
\item Operating system
\item Network (messaging) 
\end{enumerate}
The research to-date has applied the orthogonal encoded tokens to
the RDBMS, an inverted list, noSQL DB, and a message broker. I.e.
has been applied to enforce the need-to-know security of different
stack layers, that are all operating and implemented from the one
singular security policy. 

 \medskip 

To help in passing tokens securely in messages and between layers
in the stack, a straightforward modification to the token orthogonal
encoding is a follows. The approach makes interpreting token semantics
very difficult if a black-hat intercepts a message. The method is
to use a new prime number as a simple shared 'keys' between 'sender'
and 'receiver.' i.e., to obfuscate token information, that then makes
decoding a token almost impossible.  

 \medskip 

For example, the user token from section 9 above is as follows, $5*7*11*17*19=124355$.
If a prime number is selected that would never be part of the security
encoding, let's say $31$ is selected, then by taking 31 from the
encoded token $124355-31=124324$, it becomes challenging if not impossible
to factor the correct primes $\{5,7,11,17,19\}$ from the number $124324$
as long as the 'sender' and 'receiver' hold 31 as a key not to be
shared. 

 \medskip 

A similar approach can be used for the other orthogonal encoding approach
that is based on summation of indexes to some base. Indeed if the
base itself is treated as a key between 'sender and 'receiver', a
level of obfuscation is provided, as is using integer division of
the token by some prime number (prime number used as the shared key)
that is not associated with the original orthogonal sum encoding.

 \medskip 

 \medskip 

Note that by using shared keys that remove information, this increases
the information uncertainty, (higher entropy during transmission)
for the orthogonal tokens, it is much more difficult for the black-hat
to reconstruct information once removed.  

 \medskip 

These orthogonal token obfuscation techniques do not preclude the
use of formal encryption methods. However, when orthogonal token obfuscation
is used stand-alone, if performance is a significant factor that needs
resolving in terms of transmission of tokens, the approach is powerful
in protecting token semantics.

\subsection{Interoperable}

There is no one perfect security control for all technology layers
in a stack. Security controls are orchestrated and can be classified
as either 'preventative' or 'detection' controls so that when orchestrated,
they work as a whole. As such, the novel set of algorithms developed
and explained in this paper, known as orthogonal encoding, can be
made to be interoperable with other controls as required. For example,
orthogonal encoding is interoperable and can take advantage of encryption,
two-factor authentication, x500 directory services, and other known
cybersecurity controls.

\section{A complete policy implementation example}

Lets practically look at a complete policy example that has been implemented
in the RDBMS.  

 \medskip 

This trusted system policy implementation has been included to enable
the reader an understanding of the discussion above. It should be
noted the implementation described is for the RDBMS; however the security
policy outlined would be as easily implemented in the messaging layer. 

 \medskip 

 For this example, the product of the primes encoding method has been
used as the orthogonal encoding mechanism. 

 \medskip 
 
\begin{figure}[ht]
    \centering
    \includegraphics[scale=0.3]{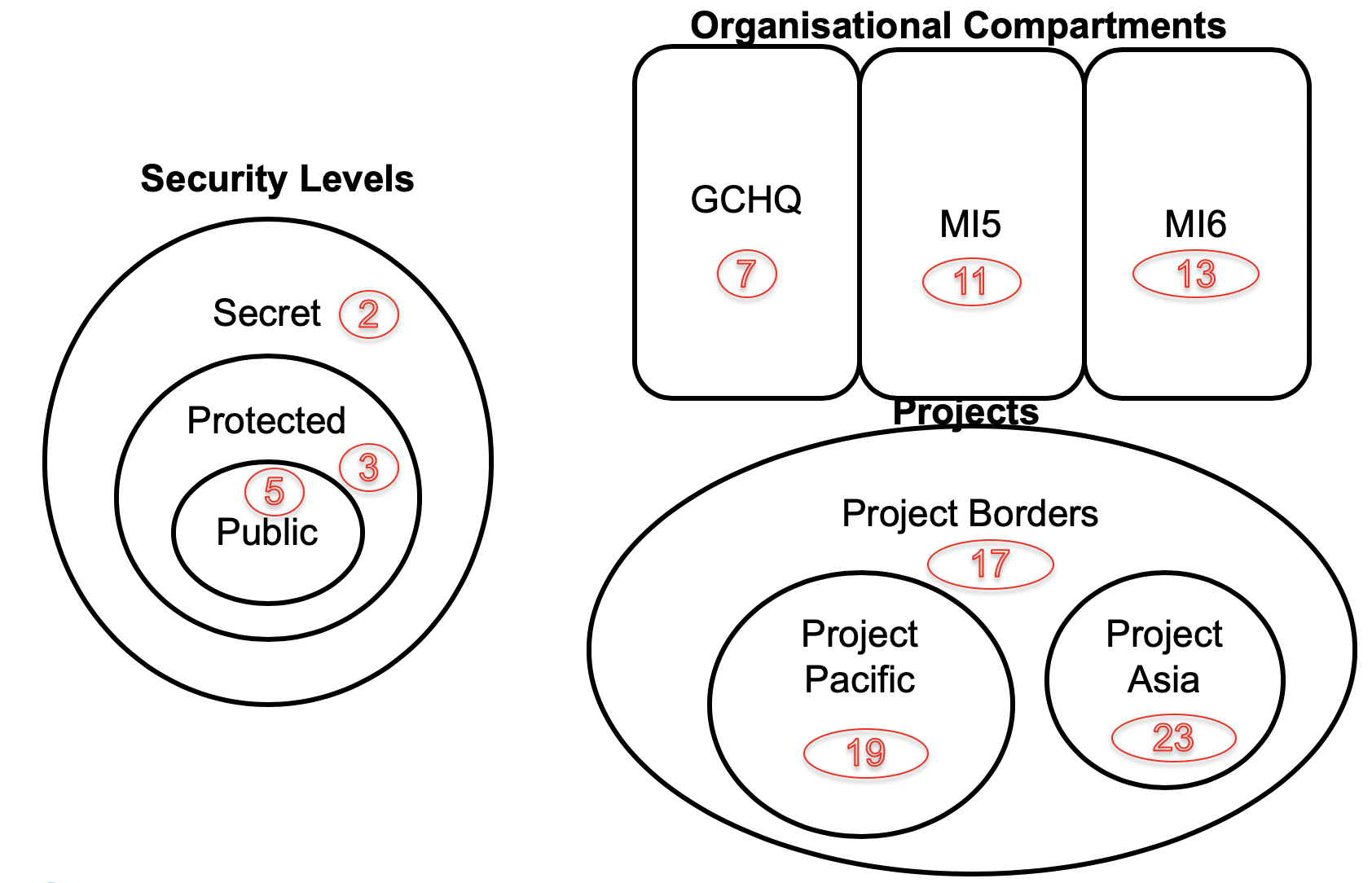} \par
    \textbf{diagram-1} - The security universe of discourse 
\end{figure}
\FloatBarrier 
 \medskip 

The second diagram indicates the user (subject) policy and the subsequent
encoding. 

 \medskip 

\FloatBarrier 
\begin{figure}[ht]
    \centering
    \includegraphics[scale=0.3]{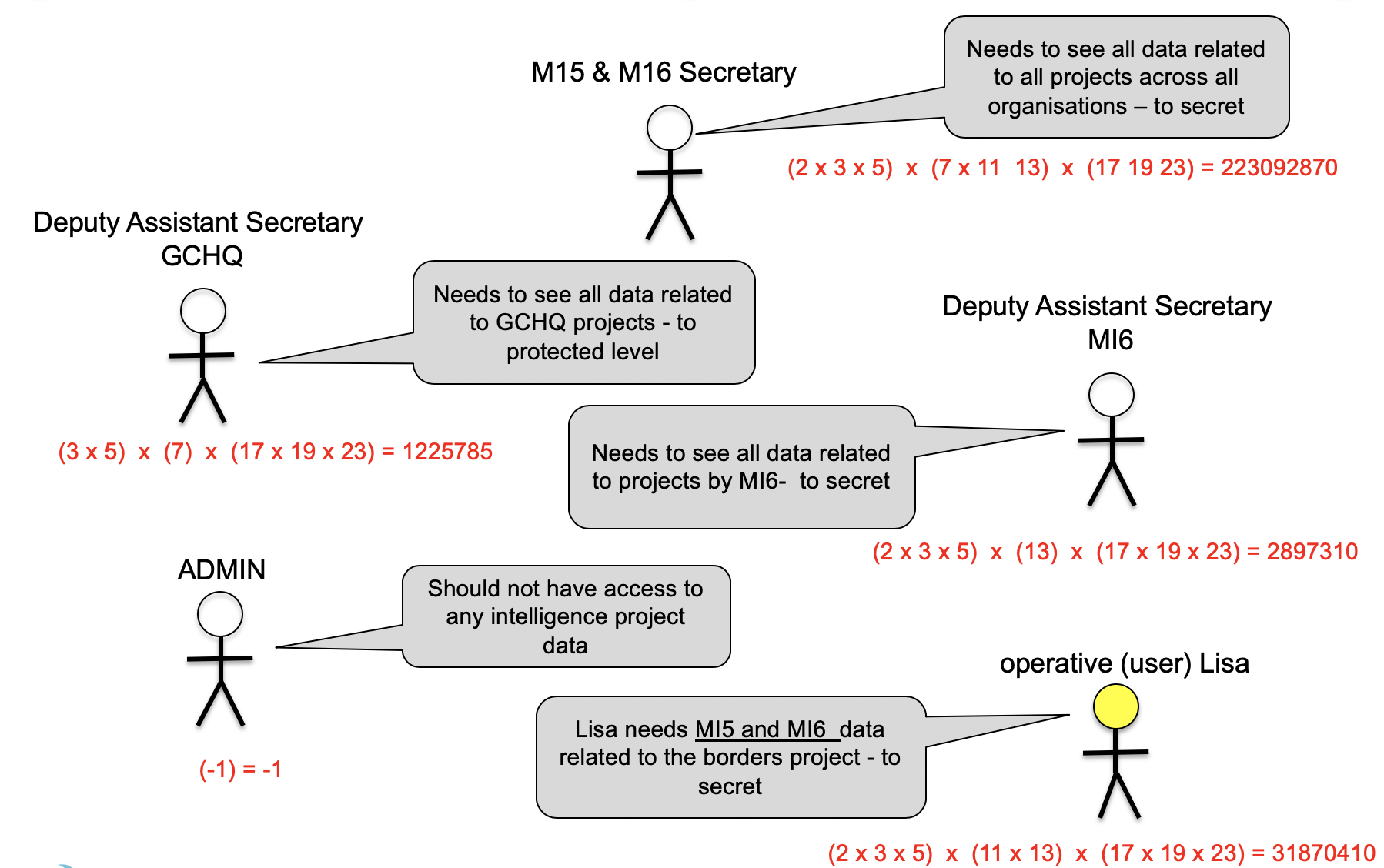} \par
    \textbf{diagram-2} - user (Subject) encoding of products of primes 
\end{figure}

 \medskip 

The third diagram indicates Security Levels and their set inclusion.
For example, if the user has access to 'Secret', they also have access
to 'Protected'. Note that set inclusion only applies to the user (Subject)
encoding 

 \medskip 
 
\FloatBarrier 
\begin{figure}[ht]
    \centering
    \includegraphics[scale=0.3]{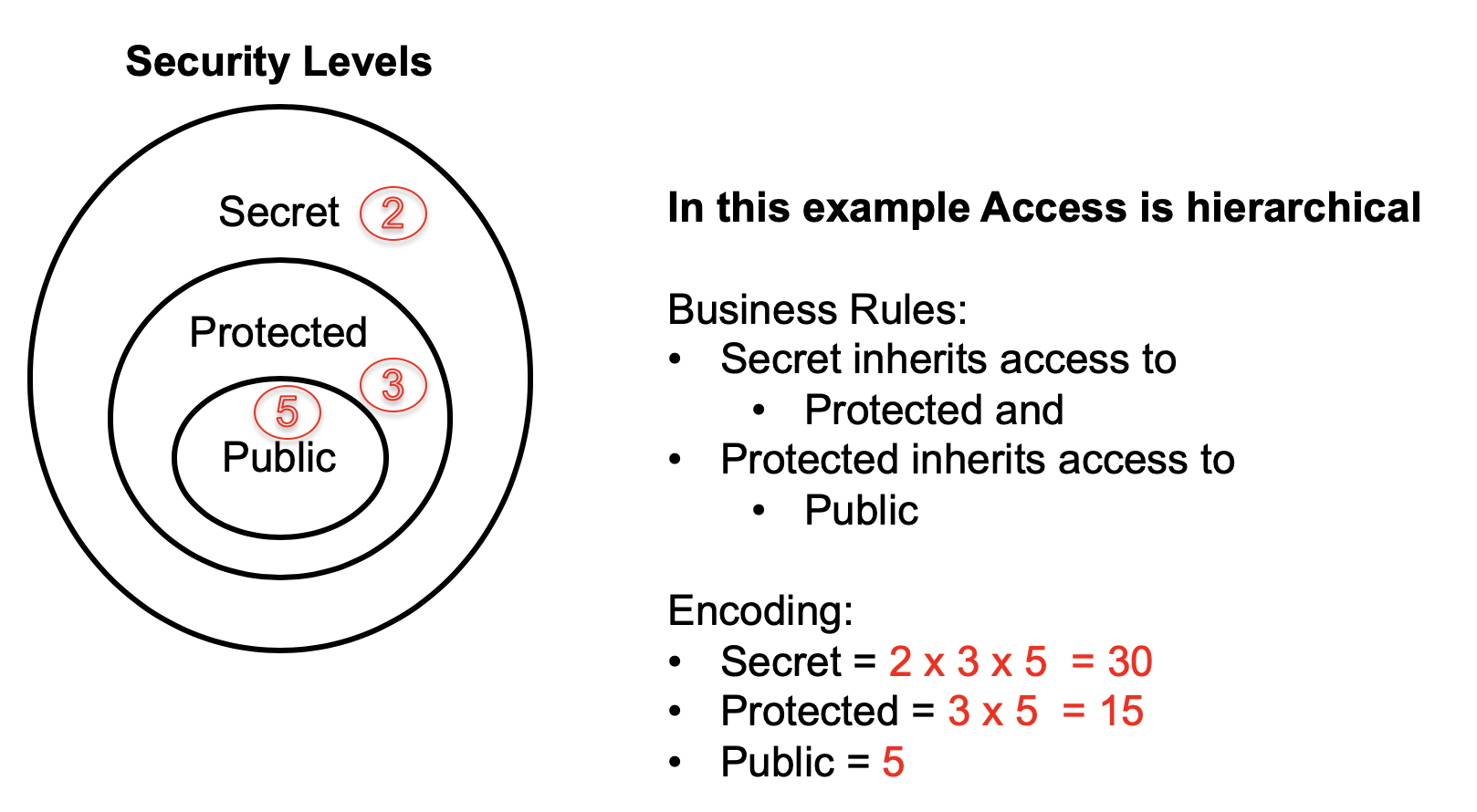} \par
    \textbf{diagram-3} - user encoding of levels is an inclusion 
\end{figure}

 \medskip 

The fourth diagram indicates the various compartments that can be
combined. There is no concept of hierarchy in terms of the compartments.
The data is just categorical.  

 \medskip 

\FloatBarrier 
\begin{figure}[ht]
    \centering
    \includegraphics[scale=0.3]{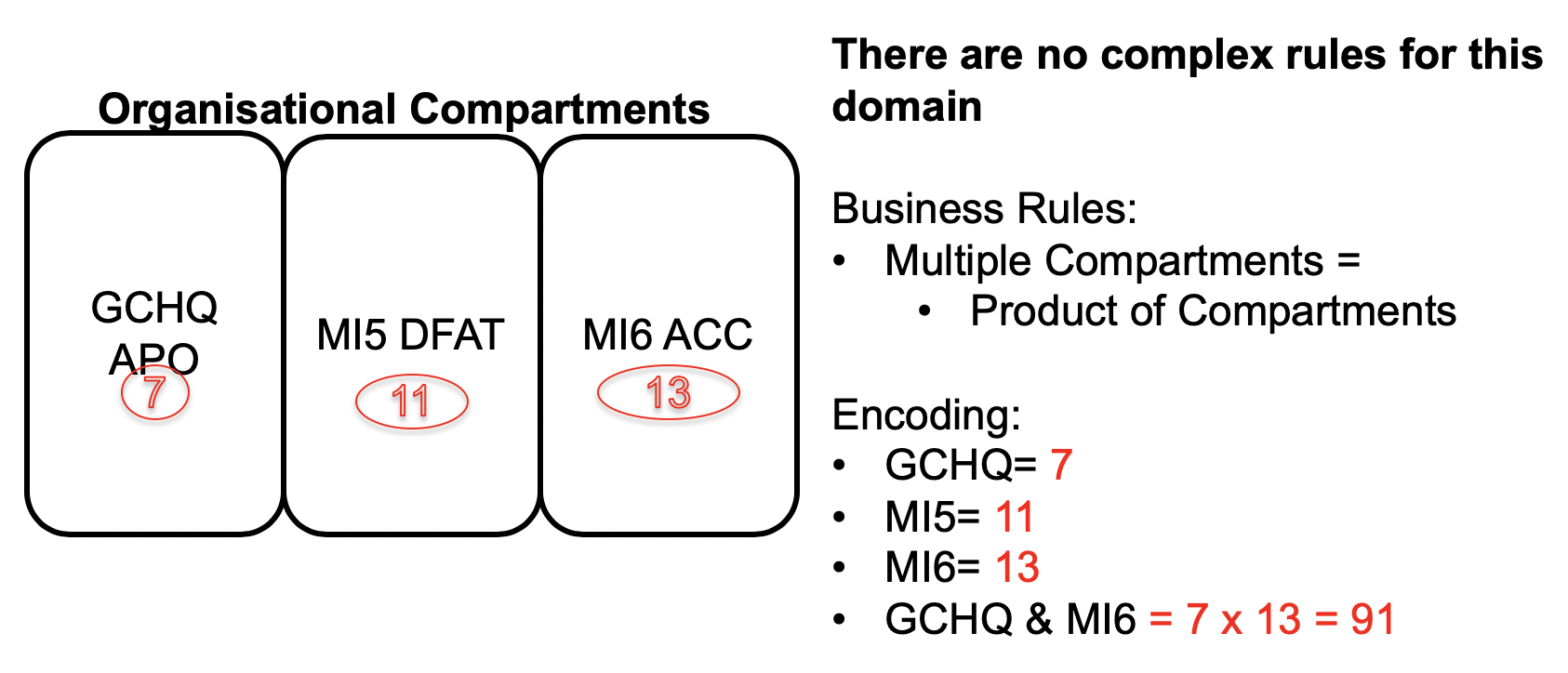} \par
    \textbf{diagram-4} - combinations of categories in this case organisational
    units
\end{figure}

 \medskip 

The fifth diagram indicates that projects are both hierarchal and
categorial. Like Security Levels, set inclusion only applies to the
user (Subject)  

 \medskip 

\FloatBarrier 
\begin{figure}[ht]
    \centering
    \includegraphics[scale=0.3]{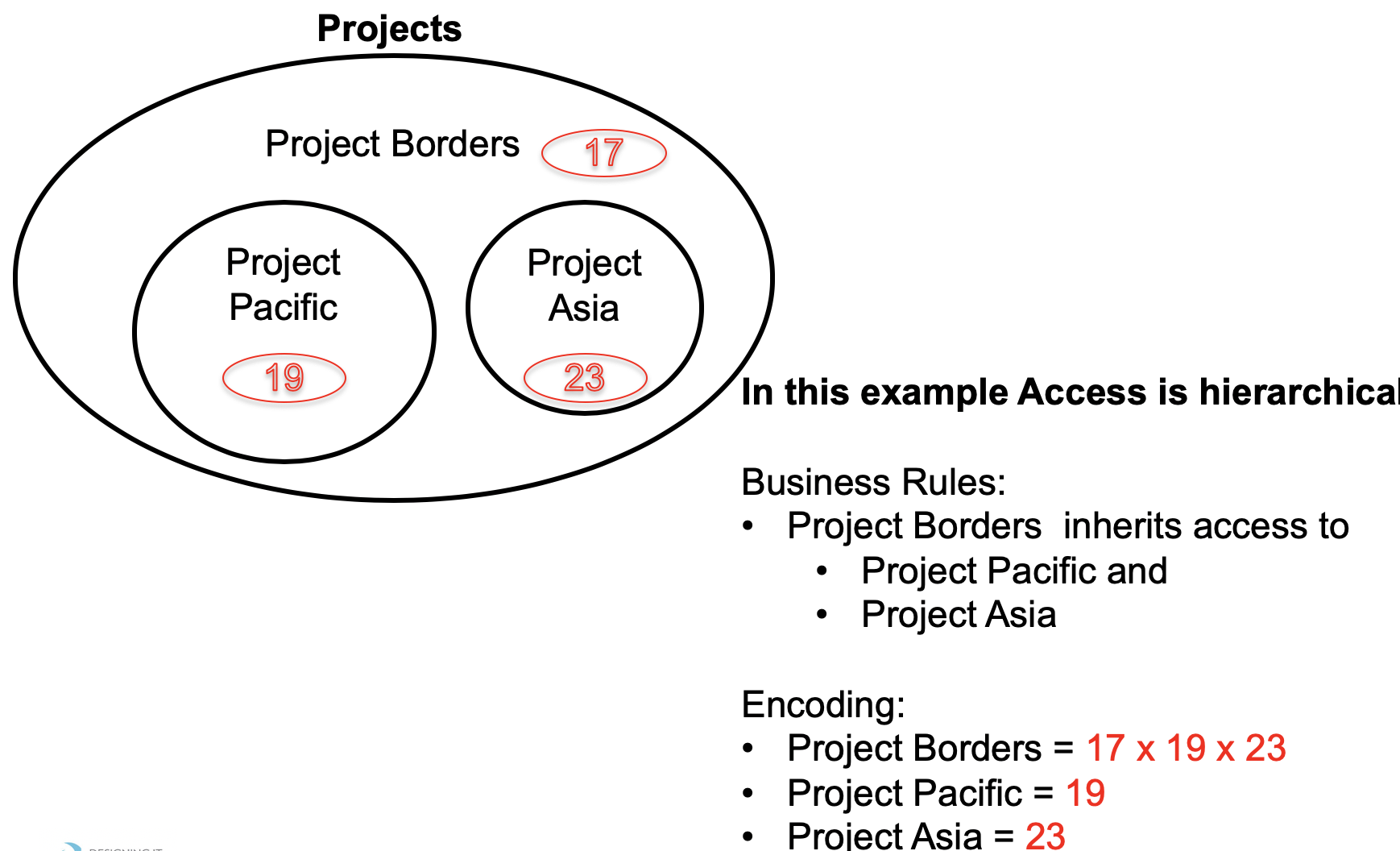} \par
    \textbf{diagram-5} user encoding of possible projects visible and
    their interrelationships
\end{figure}

 \medskip 

The sixth diagram indicates how data (Object) is encoded. Note that
Karen (the 1st row), 'borders', 'MI6' and 'Protected' are combined
as $3\times13\times17=663$ and is recorded in the 'Sec\_Tag' column.
Also note that for the data (Object) encoding, there is no concept
of set inclusion. It is not to say for projects; the argument could
not be made for set inclusion on the object. The policy, however,
treats set inclusion for projects only on the user (Subject) side. 

 \medskip 

\FloatBarrier 
\begin{figure}[ht]
    \centering
    \includegraphics[scale=0.3]{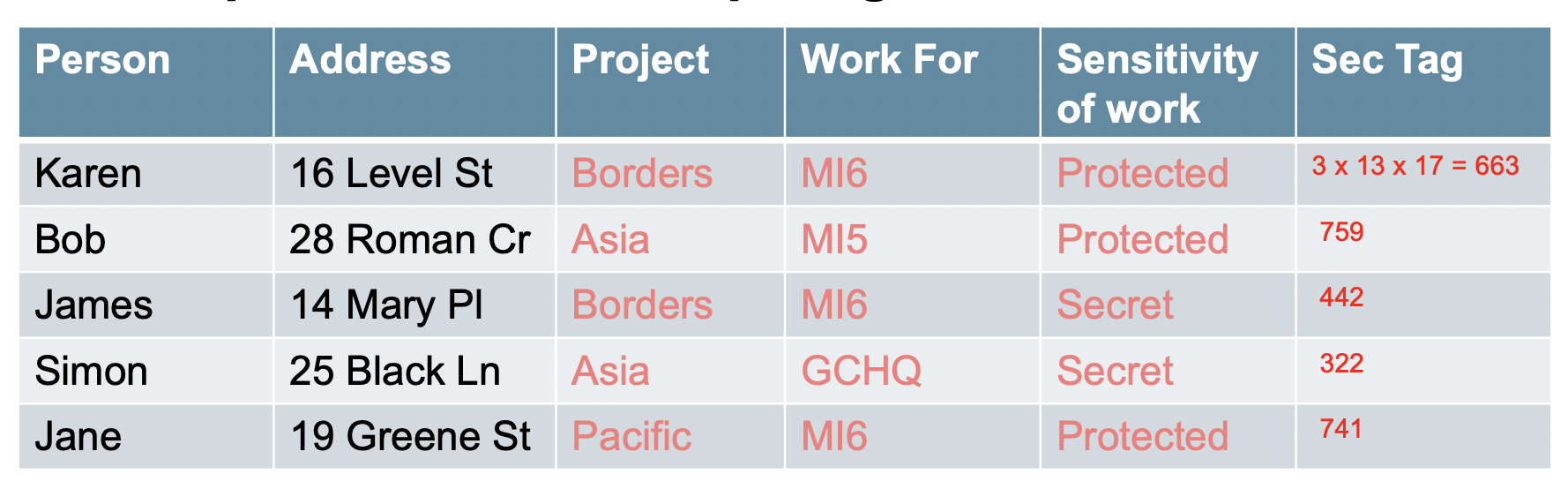} \par
    \textbf{diagram-6} - The encoding of the data (Objects) occurs at
    a row level for the RDBMS example
\end{figure}

 \medskip 

The seventh diagram indicates the mechanics of the dominance relation
at a row-level in the database. The technical implementations use
a SQL view such that the modulo function works as an abstract function
over the view based on the orthogonal encoding choices made at both
insert and update of the data. Because the modulo method is so simple
to implement, the single policy, as outlined in this example, is then
straightforward to implement in the other layers of the technology
stack, such as messaging and even 'directory services'. In this way,
the need-to-know can be defined as a single security policy, then
implemented in many technical stack layers. 

 \medskip 
 
\FloatBarrier 
\begin{figure}[ht]
    \centering
    \includegraphics[scale=0.3]{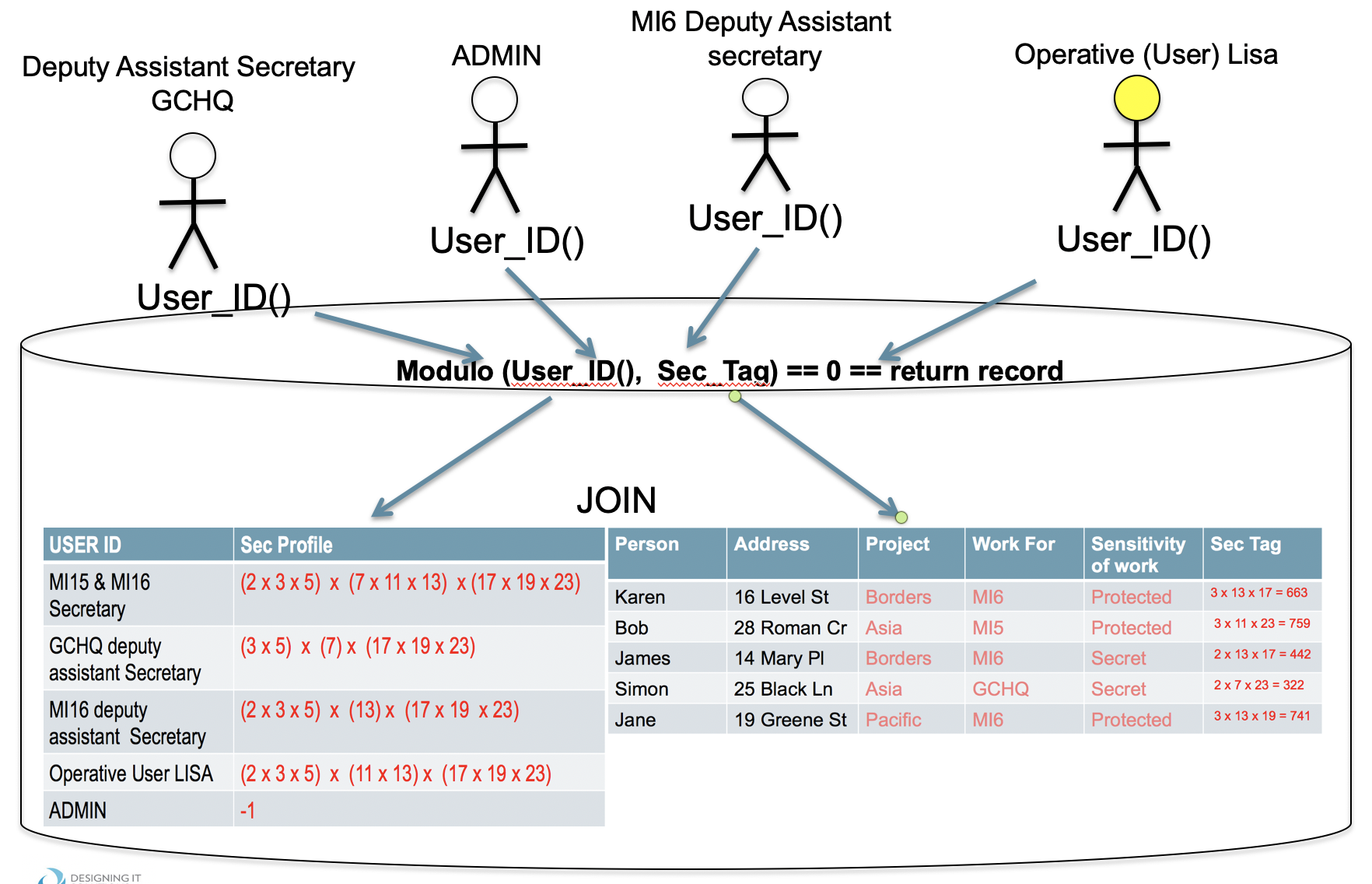} \par
    \textbf{diagram-7} - technical implementation of the dominance relation
    as a SQL view
\end{figure}

 \medskip 

Lastly, the eighth diagram indicates the rows that would be returned
(stared rows) for the MI6 Secretary. 

 \medskip 
 
\FloatBarrier 
\begin{figure}[ht]
    \centering
    \includegraphics[scale=0.3]{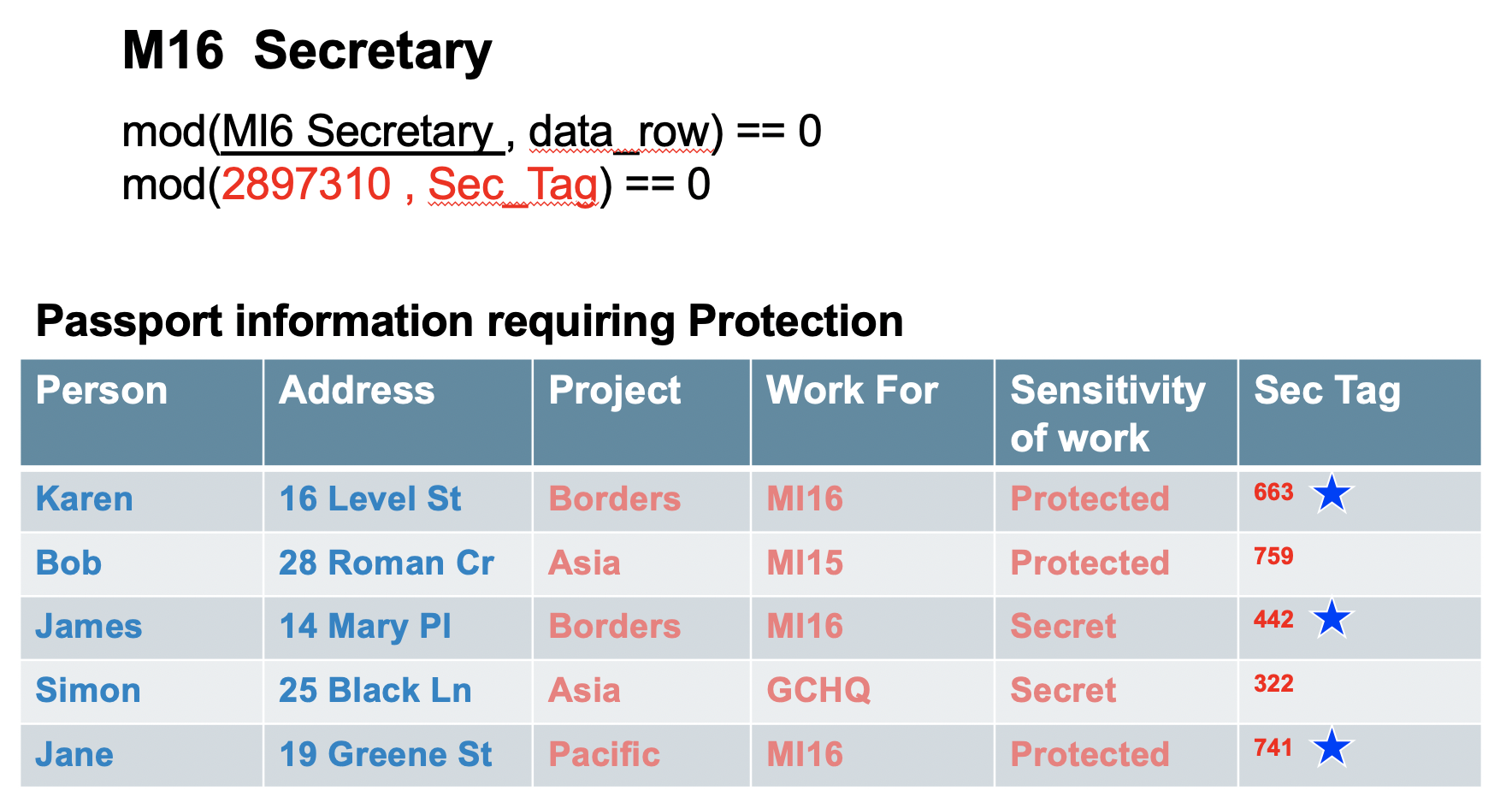} \par
    \textbf{diagram-8} - The MI6 Secretary separation of duty, their need-to-know
    on a full table scan
\end{figure}
\FloatBarrier 

\end{document}